\documentclass[structabstract]{aa}
\usepackage{natbib,twoopt}
\usepackage{graphicx,float}

\newcommand{\chandra}{{\it Chandra}}
\newcommand{\swift}{{\it Swift}}
\newcommand{\xmm}{{\it XMM-Newton}}

\newcommand{\ee}{\end{eqnarray}}
\newcommand{\be}{\begin{eqnarray}} 
\newcommand{\lp}{\left(}
\newcommand{\rp}{\right)}

\usepackage{txfonts}
\usepackage{amstext}

\begin{document}

   \title{
Short-period X-ray oscillations in super-soft novae and persistent super-soft sources
     }
   \author{J.-U. Ness\inst{\ref{esa}}\and
  A.P. Beardmore\inst{\ref{leicester}}\and
  J.P. Osborne\inst{\ref{leicester}}\and
  E. Kuulkers\inst{\ref{esa}}\and
  M. Henze\inst{\ref{esa}}\and
  A.L. Piro\inst{\ref{piro}, \ref{carnegie}}\and
  J.J. Drake\inst{\ref{sao}}\and
  A. Dobrotka\inst{\ref{slovak}}\and
  G. Schwarz\inst{\ref{aas}}\and
  S. Starrfield\inst{\ref{asu}}\and
   P. Kretschmar\inst{\ref{esa}}\and
   M. Hirsch\inst{\ref{erlangen}}\and
   J. Wilms\inst{\ref{erlangen}}
    }

\institute{Science Operations Division, Science Operations
  Department of ESA, ESAC, Villanueva de la Ca\~nada (Madrid), Spain; corresponding author:
  \email{juness@sciops.esa.int}\label{esa}
\and
Department of Physics \& Astronomy, University of Leicester, Leicester, LE1 7RH, UK\label{leicester}
        \and
Theoretical Astrophysics, California Institute of Technology, 1200 E California Blvd, M/C 350-17, Pasadena, CA 91125, USA\label{piro}
     \and
Observatories of the Carnegie Institution of Science, 813 Santa Barbara Street, Pasadena, CA 91101, USA\label{carnegie}
     \and
Harvard-Smithsonian Center for Astrophysics, 60 Garden Street, Cambridge, MA 02138, USA\label{sao}
        \and
Advanced Technologies Research Institute, Slovak University of Technology in Bratislava, Paulinska 16, 91724 Trnava, Slovak Republic\label{slovak}
        \and
American Astronomical Society, 2000 Florida Ave., NW, Suite 400,
DC 20009-1231, USA\label{aas}
\and
School of Earth and Space Exploration, Arizona State University, Tempe, AZ 85287-1404, USA\label{asu}
        \and
Remeis Sternwarte \& Erlangen Centre for Astroparticle Physics, Sternwartstr. 7, 96049, Bamberg, Germany\label{erlangen}
}
   \authorrunning{Ness et al.}
   \titlerunning{Short-period X-ray oscillations in SSS}
   \date{Received \today; accepted }

  \abstract
% Context
{
%Super Soft Sources (SSS) emit atmospheric continuum X-ray radiation
%with an effective temperature of $10-100$\,eV, powered by nuclear
%burning on the surface of a white dwarf.
%They are commonly assumed to be powered by nuclear burning on the
%surface of a white dwarf, requiring a nearby companion star from which
%the hydrogen-rich fuel is obtained via accretion. In addition to
%persistent SSS, novae go through a phase of SSS emission after the
%ejecta have cleared.
Transient short-period ($<100$\,s) oscillations have been
found in the X-ray light curves of three novae during their super-soft
source (SSS) phase
and in one persistent SSS.
% The origin of these transient short-period oscillations is unknown.
}
{
We pursue an observational approach to determine possible
driving mechanisms and relations to fundamental system
parameters such as the white dwarf mass.
}
{
We performed a systematic search for short-period oscillations
in all available \xmm\thanks{\xmm\ is an ESA science mission with
   instruments and
   contributions directly funded by ESA Member States and NASA.}
and \chandra\thanks{Software provided by the \chandra\ X-ray Center
   (CXC) in the application package CIAO
   was used to obtain science data.}
X-ray light curves of persistent SSS and novae during their SSS
phase. To study time evolution, we divided each light curve into
short time-segments and computed power spectra.
We then constructed a dynamic power spectrum from which we
identified transient periodic signals even when only present
for a short time. We base our confidence levels on simulations
of false-alarm probability for the chosen oversampling rate of
16, corrected for multiple testing based on the number of
time segments. From all time segments of each system, we
computed fractions of time when periodic signals were detected.
   }
   {
In addition to the previously known systems with short-period 
oscillations, RS\,Oph (35\,s), KT\,Eri (35\,s), V339\,Del
(54\,s), and Cal\,83 (67\,s), we found one additional system,
LMC\,2009a (33\,s), and also confirm the 35\,s period from
\chandra\ data of KT\,Eri. The oscillation amplitudes are
of about $<15$\% of the respective count rates and vary without
any clear dependence on the X-ray count rate.
The fractions of the time when the respective periods
were detected at $2\sigma$ significance (duty cycle) are
11.3\%, 38.8\%, 16.9\%, 49.2\%, and 18.7\% for
LMC\,2009a, RS\,Oph, KT\,Eri, V339\,Del, and Cal\,83,
respectively. The respective highest duty cycles found in a
single observation are 38.1\%, 74.5\%, 61.4\%, 67.8\%, and 61.8\%.
   }
   {
Since fast rotation periods of the white dwarfs as origin of
these transient oscillations
are speculative, we concentrate on pulsation mechanisms.
We present initial considerations
predicting the oscillation period to scale linearly with the
white dwarf radius (and thus mass), weakly with the pressure
at the base, and luminosity. Estimates of the size of the
white dwarf could be useful for determining whether these
systems are more massive than typical white dwarfs, and thus
whether they are growing from accretion over time. Signs of
such mass growth may have implications for whether some of
these systems are attractive as Type Ia supernova progenitors.
    }

   \maketitle
%
%________________________________________________________________

\section{Introduction}

The class of super-soft sources (SSS) was empirically
defined as X-ray sources emitting a blackbody-like spectrum yielding
temperatures between 20--100\,eV (3--7$\times 10^5$\,K) and
luminosities above $10^{35}$\,erg\,s$^{-1}$ \citep{greiner96}.
The first such sources were found with the {\it Einstein}
Observatory by \cite{cal_discovery} and were later defined as
a class after more were discovered with ROSAT
\citep{truemper92,sssclass,sssclass1}. It is now commonly
accepted that SSS emission originates in binary systems
containing a white dwarf primary that hosts nuclear burning of
material that is accreted from a secondary star; see
\cite{heuvel} and \cite{kahab}. Based on this interpretation,
the term SSB (super-soft X-ray binary) is sometimes used.
In high spectral resolution data that can be achieved with the gratings
on board \xmm\ and \chandra, details can be seen that clearly
confirm theoretical expectations that the X-ray spectra are
not a blackbody (as e.g. several neutron stars), but an atmo\-spheric
continuum spectrum with absorption lines and edges. Important
system parameters can be derived using atmsophere models, which
was, for example, demonstrated for the famous persistent SSS Cal\,83 by
\cite{lanz04}.\\

In addition to the persistent SSS, classical and
recurrent novae (CNe, RNe) pass through a phase during which they
emit an SSS spectrum in X-rays that is thought to originate from the
hottest
layers closest to the surface of the white dwarf. The SSS phase lasts
until the hydrogen content of the accreted material is consumed
or ejected. The X-ray and UV evolution of
novae during their outbursts has been determined using \swift\
monitoring observations. Deeper, continuous \xmm\ and \chandra\
observations have been obtained, guided by the long-term
evolution of UV/X-ray emission determined by the \swift\ UVOT
and XRT instruments.
% Several but not all novae have shown early hard emission that
%might come from shocks within the ejecta or with the surrounding
%medium.
The turn-on time of the SSS phase, when bright SSS emission becomes
visible, depends on the evolution of the nova ejecta. Accurate
predictions are currently not possible from evolutionary models, but
empirical scaling relations have recently been derived from population
studies \citep{henze13,schwarz2011}. Their predictive power is not
yet well established, and \swift\ is still needed to study the
X-ray evolution of individual novae and to guide deeper X-ray
observations.\\

The first \swift\ X-ray monitoring campaign of a nova was performed
during the 2006 outburst of the recurrent nova RS\,Oph, and
\cite{atel770} found short-period oscillations of about 35\,s
in some but not all X-ray observations taken during the SSS phase.
This transient signal in RS\,Oph was studied in more
detail by \cite{osborne11}. The period was first seen on day
32.9 after outburst and last detected on day 58.8, consistent
with the start of the SSS phase and the onset of the decline,
respectively. Independent evidence for this
period was reported by \cite{ness_rsoph} in an \xmm\ observation
taken on day 54 after outburst. They confirmed that the signal came
from the SSS component and not from the shock emission that was
present in the same observation at higher energies.
\cite{nelson07} also reported having seen this signal for a
short time in an \xmm\ observation taken on day 26.1.\\

Surprisingly, a transient 35s period was also seen in
\swift/XRT data obtained for KT\,Eri \citep{kteri35}.
Most recently, a 54s transient period was found for the nova
V339\,Del in \swift/XRT \citep{v339del_35} and \xmm\ observations
\citep{ness_v339del}, both taken during the SSS phase. 
Two \chandra\ observations taken on 2013 November 9 (day 87.2)
and 2013 December 6 (day 114) also contain the 54s period (Nelson
private communication, publication in preparation).\\

Similar X-ray periods were found in probably non-burning white
dwarfs, for instance,
a 33s spin of AE Aqr \citep{aeaqr33}, $\sim 29$\,s
in WZ Sge \citep[][and references therein]{nucita14},
or few-second quasi-periodic oscillations (QPOs) in a number of dwarf novae,
e.g., SS Cyg, U Gem, \citep{cordova84}; referred to as
dwarf nova oscillations, DNOs, see for example,
\cite{maucheqpo,warner03} and references therein.
Since the X-ray spectra of these systems reflect distinctly
different production mechanisms from SSS spectra (e.g.,
accretion disc or the boundary layer), we do not include
these CV systems in our study.\\ 

In addition to novae, a period of $\sim 67$\,s
has been seen by \cite{cal83_67sec} in the prominent
persistent SSS Cal\,83, which they interpreted as driven by the
rotation period of the white dwarf. While this is close to
the break-up period, they argued that the white dwarf may be
spun up by accretion disc torques. A $\pm3$\,s drift from the median value in this 
scenario would be more difficult to explain (see the discussion in Sect.~\ref{disc:rot}).\\

All indicators suggest that the short-period oscillations
originate in the SSS component.
% While other evolutionary
%phases of novae are fainter in X-rays, some non-SSS
%observations are bright enough to detect such signals,
%e.g., the shock component in RS\,Oph \citep{ness_rsoph}.
This type of short-period oscillation may thus originate
from the surface of the white dwarf while undergoing nuclear
burning. It will be of interest to determine whether
the 35--54s periods in the novae are related to the
67s period in the persistent SSS Cal\,83. If the origin is
not the spin period of the white dwarf, it might be related
to the interiors or the nuclear burning regions, and
understanding these processes would give a new diagnostic
method to determine fundamental parameters such as the mass
of the underlying white dwarf. It is therefore important
to find as many systems as possible that host short-period
oscillations, ideally covering a wide range of system
parameters, to identify those properties that
drive them.\\

We have searched for short-period variations in all systems
showing bright SSS emission, focusing here on \xmm\ and
\chandra\ observations, while a similar project based on all
\swift/XRT data will be presented by Beardmore et al. For systems with short-period oscillations, we study the
evolution of the power and the period.\\

We briefly describe the techniques and the observations used in this
article in Sect.~\ref{observations}. We then describe the timing analysis
in Sect.~\ref{sect:analysis} and the results of it in
Sect.~\ref{sect:results}.
% where we dedicate an extra
%subsection to each system in which short-term oscillations
%are present, closing with Sect.~\ref{sect:other} in which
%we briefly describe those observations in which no periodic
%signal was found.
We discuss our results in Sect.~\ref{sect:disc} and summarise
our findings and conclusions in Sect.~\ref{sect:concl}.

\section{Observations}
\label{observations}

%X-ray photons of astronomical sources are focused with nested
%mirror shells facilitating shallow reflection angles to avoid
%X-rays to be absorbed in the mirrors. The detectors are placed
%in the focal plane. \chandra\ uses a sliding mechanism
%to move a desired instrument into the focal plane of a single
%mirror set. To minimise the risk of failure of moving parts,
%\xmm\ has three sets of mirrors with different instrument
%configurations behind each of them, where only the filter
%wheels are moving parts. \xmm\ also operates an optical/UV
%monitor (OM) in parallel.\\
%The X-rays are recorded with CCD detectors measuring photon
%positions on the chip, arrival times, and photon energies.
%\chandra\ also has a microchannel plate for achieving high
%resolution imaging and timing, albeit without energy information.
%In addition, transmission or reflecting gratings can be used
%to disperse the incident light before it is recorded by the
%detectors, yielding high-resolution spectra. 

\xmm\ and \chandra\ observe from highly elliptical orbits
allowing long uninterrupted observations for up to two days. The
instrumentation on board provides high resolution in time
and energy using various combinations of CCD detectors and
dispersive gratings. For more details, we refer to the
corresponding papers about the \xmm\ satellite \citep{xmm}
hosting the  European Photon Imaging Camera, EPIC,
consisting of the pn detector \citep{epic_pn} and the
Metal Oxide Semi-conductor, MOS, \citep{epic_mos}, and the
Reflection Grating Spectrometers, RGS, \citep{rgs}.
The \chandra\ satellite is described by \cite{chandra}
with the High Resolution Camera HRC, \citep{hrc}, the
Low Energy Transmission Grating, LETG, \citep{letg}, and
the Advanced CCD Imaging Spectrometer, ACIS \citep{acis}.\\

A large sample of high-resolution X-ray grating spectra of
SSS was presented by \cite{ness_obsc}, where
a list of observations is given in their Table~2, from
which we selected the observations in this work. We
used the light curves from the contemporaneous \xmm/EPIC
detectors or zero order for \chandra\ transmission
grating observations.
In Tables~\ref{tab:obs} and \ref{tab:obs2} we list the
target names, day after outburst for novae (reference
day given in Table~1 of \citealt{ness_obsc}), observation
identifier, detector from which light curves were extracted
(see table footnotes and text below), start date, exposure time,
and the bin size used for the period analysis. The following
five columns give the larger bin size used to determine count rates given in the following four columns, minimum
and maximum count rates (thus the range of variability)
supplemented by median and mean count rates. These numbers
give an idea of the brightness and the degree of variability.
In the last column we list a  code that indicates the detrending
method (see Sect.~\ref{sect:analysis}).\\

%For this work we analyse the X-ray light curves obtained with
%the \xmm\ EPIC and RGS and with the \chandra\ HRC and ACIS
%obtained by filtering the events on source and background
%positions plus focusing on soft energies and binning the
%photon arrival times into a chosen time grid.\\

%\input{obstab}

\begin{table*}
\begin{flushleft}
\renewcommand{\arraystretch}{1.1}
\caption{\label{tab:obs}Journal of X-ray observations of super-soft X-ray sources}
{%\scriptsize
%\begin{tabular}{lllrlrrrrrrrl}
%\begin{tabular}{lllrlrp{.1cm}rp{.2cm}rrrl}
\begin{tabular}{lllrlrp{.1cm}rp{.7cm}p{.7cm}p{.7cm}p{.7cm}l}
\hline
Target & Day$^a$ & \multicolumn{2}{l}{ObsID$^b$  \hfill Detector$^c$} & \multicolumn{2}{l}{Start time \hfill Exp.\,time$^d$}&\multicolumn{2}{c}{bin size}&\multicolumn{2}{l}{Min.\ \ Median}&Mean&Max.&D-mode$^g$\\
&&&&(UT)&(ks)&(s)$^e$&(s)$^f$&\multicolumn{4}{c}{count rate (counts s$^{-1}$)}&\\
\hline
\input{obstab.lines}
\end{tabular}
}

$^a$After $t_{\rm ref}$ (see Table~1 in \citealt{ness_obsc})\\
$^b$Observation identifiers\\
$^c$On board \xmm\ (pn, MOS, RGS) and \chandra\ (HRC, ACIS). For pn, mode and filter are given as FF=full frame, LW=large window, SW=small window, Ti=timing, and tn=thin filter and m=medium filter. \\
$^d$Exposure time (ks$=10^3$\,s)\\
$^e$Time bin size for analysis\\
$^f$Time bin size for determination of max/min count rates\\
$^g$Detrending mode, see text.\\
$^h$Reference date for V339\,Del: 2013-8-14.584

\renewcommand{\arraystretch}{1}
\end{flushleft}
\end{table*}

For \xmm\ observations, we extracted light curves from the EPIC
detectors pn in 1s time bins.
V4743\,Sgr, RS\,Oph, and V339\,Del were too bright for the
EPIC cameras, which suffered excessive event losses from
buffer overflow. For these novae, we used the light curves
from either the less sensitive MOS cameras (V339\,Del)
or from the Reflection Grating Spectrometers (RGS).
The instrument with mode and filter used is given in the
fourth column of Table~\ref{tab:obs}. Some details of the
instruments are also included in the corresponding
graphical representations of the results.\\

The raw data were processed with standard SAS tools
of version 13.5. We started with a standard run using the
general SAS tool {\tt xmmextractor,} which delivers a full
set of event files and science products using standard
parameter settings, some of them optimised to the specific
observation. We then inspected the science products and
re-generated the EPIC and RGS light curves over the
0.1--10-keV range using {\tt evselect} and {\tt epiclccorr}
or {\tt rgslccorr}. We furthermore extracted all available
light curves that were taken with the optical monitor in
fast mode in time binnings of 0.5 seconds.\\
%with and without absolute corrections.

All \chandra\ observations in our sample used the
LETG. We extracted the light curves from the zero-order
photons of the HRC or the ACIS (Cal\,87) using the regions
of zero order that are automatically created during
extraction of grating spectra. The extraction was made
with the task Ciao v. 4.6  {\tt dmextract} in binning of
1\,s (for HRC) and 2\,s for ACIS, without energy filtering.

\section{Analysis}
\label{sect:analysis}

Before searching for periodic oscillations, we detrended
the light curves. We first fitted either a high-order polynomial
to the light curve or smoothed it with a Gaussian kernel. The
polynomial or smoothed curves were then subtracted from the
light curve and the difference used for further analysis.
In the last column of Table~\ref{tab:obs} we list the detrending
mode as P$n$ when using an $n$-th order polynomial and G$n$
when using a Gaussian-smoothed light curve, with $n$ reporting the
full width at half maximum (in seconds).\\

The period searches were performed using power spectra calculated
with the method of \cite{hornebal}. The details are described
in the appendix. Since past observations imply that we here investigate a transient signal, we have to consider the possibility that
a signal might not be detectable in a full light-curve. To detect any sporadic appearance within an observation,
we therefore computed power spectra from overlapping $t=1000$s time
segments in adjacent steps of 500\,s. In all cases we tested
$N_{\rm f}=500$ frequencies within a range of periods between
25\,s and 100\,s. This range covers all periods so far
reported and avoids contamination by low-frequency noise.\\

In the Appendix we present simulations supporting the way we
interpret the resulting powers from the power spectra as
likelihood of detecting a periodic signal. We also discuss
corrections for multiple testing when studying various time
intervals from a single light-curve based on the conservative
Bonferroni correction \citep{bonf36,dunn59,dunn61}.\\

\begin{figure*}[!ht]
\resizebox{\hsize}{!}{\includegraphics{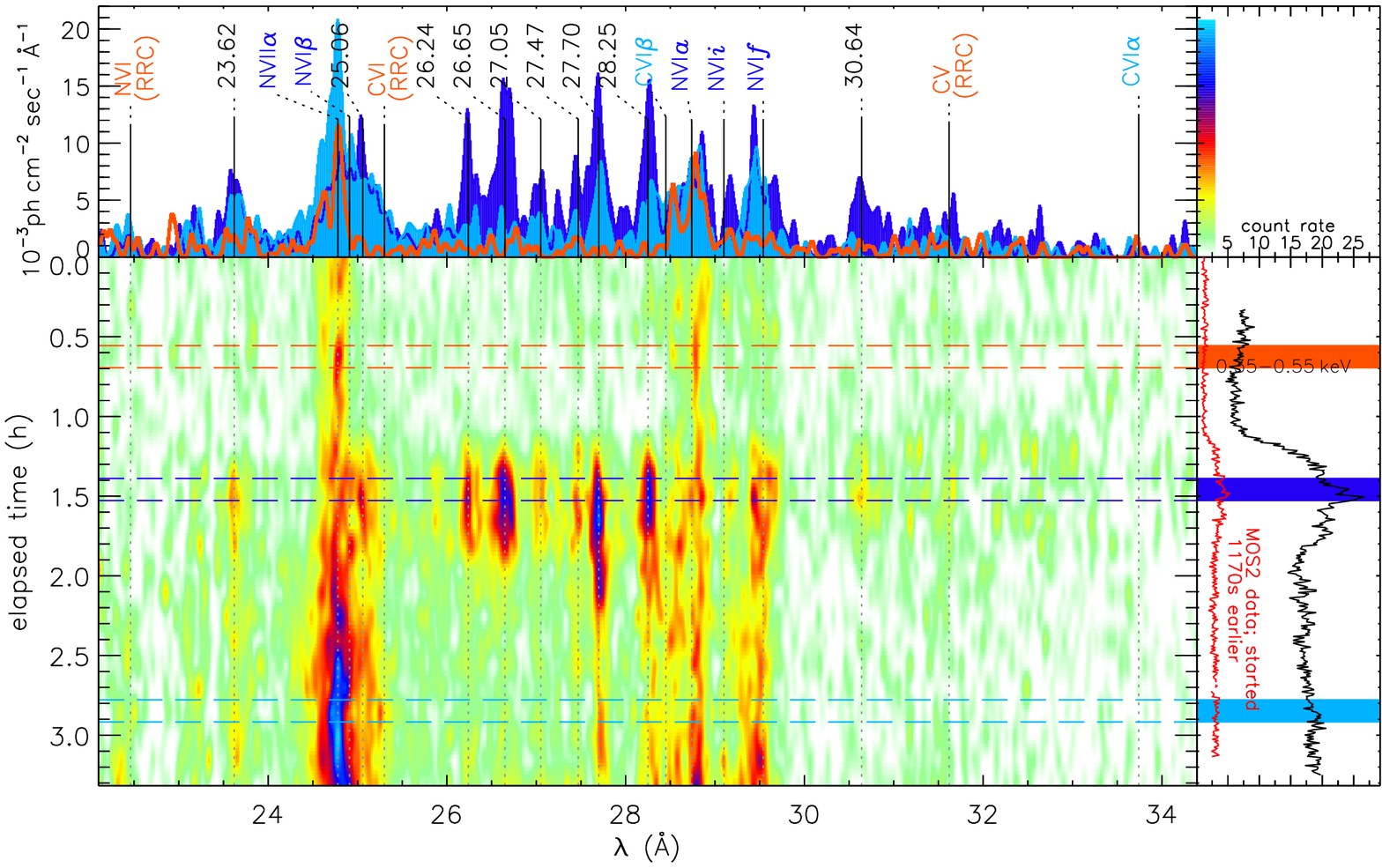}\includegraphics{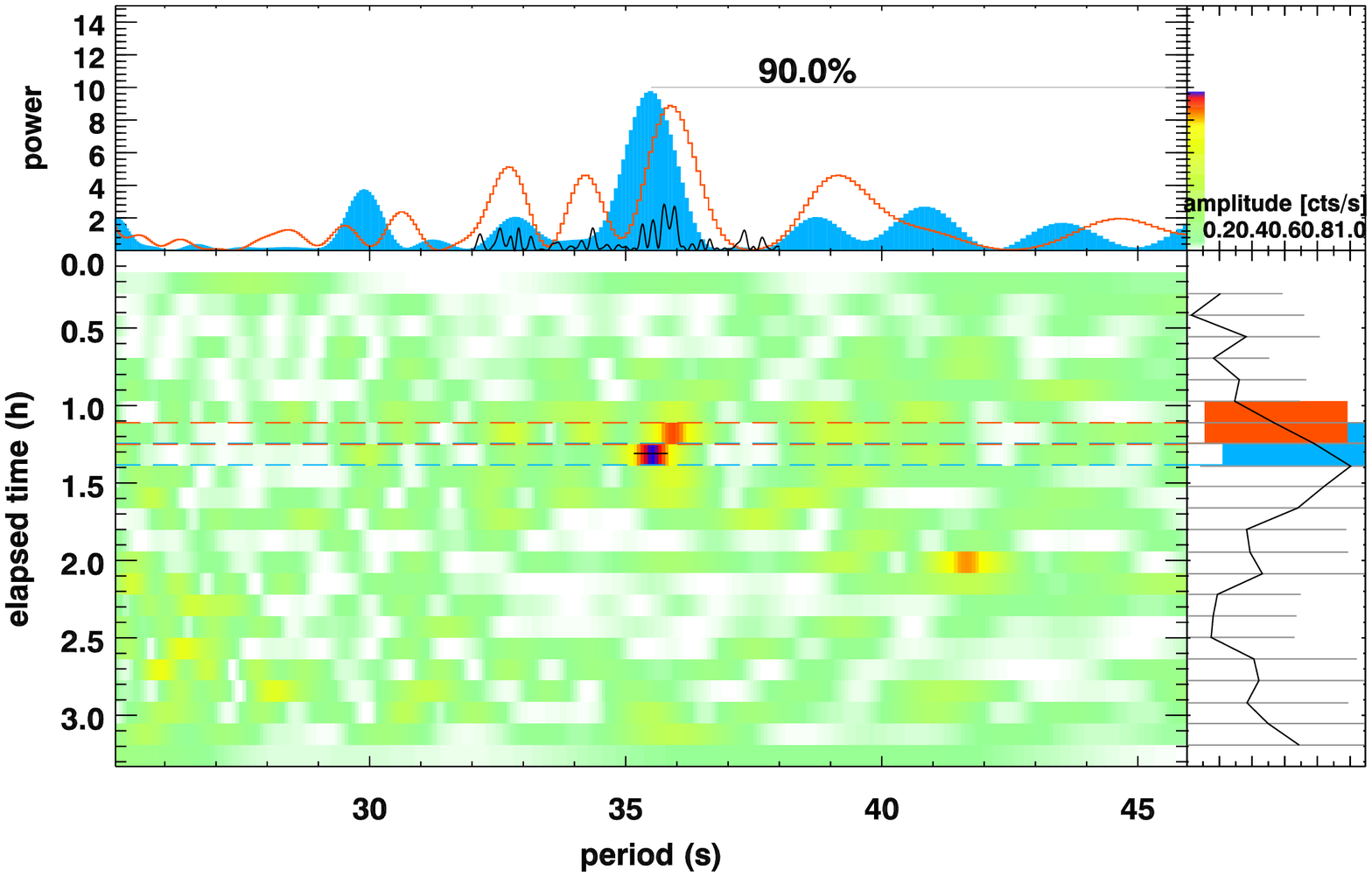}}
\caption{\label{smap_lmap}RS\,Oph on day 26.1: Spectral (left) and
period (right) time-maps on the same time scales in vertical
direction.
{\bf Left}: Spectral time-map showing the evolution of the RGS
spectrum; taken from \cite{ness_obsc}, adding unidentified emission
features with their wavelengths in \AA. The simultaneous EPIC/pn
and MOS2 light curves are shown in the right panel, rotated clockwise
by 90$^{\rm o}$. The pn has a longer initial overhead such that the
exposure started $\sim 1170$\,s later than the RGS and the MOS exposures,
thus the black curve (pn) starts at 0.33\,h, while the red curve (MOS)
starts at 0\,h.
{\bf Right}: Dynamic power spectrum consisting of the power spectra
computed from 24 overlapping 1000s time segments from the pn light curve.
To map them into
the 11ks time frame of the observation, the width of each row in the
central panel is reduced to 500s. The horizontal orange/blue dashed
lines running across the main panel border two adjacent 500s time intervals
that correspond to the (wider) coloured bars in the right panel.
They mark
the overlapping 1000s time intervals from which the individual
power spectra were computed that are shown with the same
colours in the top panel. The colour scheme for the main panel is
decoded in the top right part; blue colours were chosen to mark
periods with $>$90\% confidence and red for 30--80\%,
while yellow/light green reflect noise.
Colour-coding and thresholds have been corrected for oversampling and
multiple testing (see text).
The short horizontal black line in the main panel at 35s represents
the 90\% uncertainty range in period of a significant ($>2\sigma$) signal.
In the top panel, in addition to the two power spectra from partial
light-curves, the power spectrum from the full light-curve
is shown in black; the same oversampling factor of 16.67 has been
applied and contains no significant signal.
The widths of the peaks in the respective power spectra are consistent
with the intrinsic period resolution imposed by windowing.
At the right, we show the evolution of the modulation amplitude,
most of them are upper limits.
}
\end{figure*}

To illustrate the evolution of a periodic signal, we display the
series of power spectra with the concept of a dynamic power spectrum,
similar to the dynamic spectra that were used for X-ray
spectra by \cite{basi}, for example. A comparison is shown in
Fig.~\ref{smap_lmap} (discussed in Sect.~\ref{sect:oph}), where
time evolution maps of the X-ray
spectrum (left panel) and of the power spectrum (right panel)
are shown. The light curve is shown with time running down in
the right part of the left panel, turned around by 90$^{\rm o}$
clockwise, where in this case both EPIC/pn and EPIC/MOS light
curves of RS\,Oph are shown in units of count rates. In the
right plot, the modulation amplitudes are shown
in units of counts per second
for each 500\,s time bin along the same vertical time axis. The
horizontal period axis in the right plot is centred on the 35\,s
period found in the later \swift\ and \xmm\ observations. While we
always searched for periods over 25--100\,s, we here show only the
period range around the significant signal. We defined the colour
scheme based on detection likelihood using red at
$\sim$ 30--80\% and blue at $>90$\%. The colour scheme and the
marked confidence levels are based on the thresholds after
correcting for multiple testing and oversampling.
The colour-encoding can be identified from the colour bar and
the corresponding horizontal lines in the upper right corner.\\

We first searched for significant signals within the entire
25--100s period interval in each time segment. For the strongest
peak, we focused on a narrow period interval $\pm 2$\,s to
determine the period from a spline fit. We then determined
the amplitude from a sine fit to the folded light curve.
Rough uncertainties in amplitude were estimated by observing the increases
in $\chi^2$ when varying the amplitude from the best-fit value
while keeping the other two parameters (period and phase) fixed
at their respective best-fit values.\\

We then calculated the uncertainty of the pulse period in two
different ways. In the first approach, we computed the standard
deviation using Eq. (4) in \cite{larsson96}, using the
standard deviation of count rates in the corresponding
time segment, $\sigma_\mathrm{tot}$, and the amplitude of the best-fit sinusoid fitted to the folded light curve; see also Eq.
(14) in \cite{hornebal} and \cite{kovacs81}. Here we neglected the
uncertainties in the amplitude. In the second approach, we
determined the standard deviation of the pulse periods obtained
from Monte
Carlo simulated light curves. These light curves were obtained by
replicating the folded light curve using the same sampling as the
original light curve and adding the appropriate Poisson noise.
We then calculated the standard deviation of the periods measured
from 10,000 such realizations and assigned this number to the
uncertainty of the period. In general, both approaches yield
comparable results. In this work we use twice the standard
deviation as estimates of uncertainties in period. Assuming
Gaussian statistics, this would correspond to 95\% confidence
ranges. The respective results are illustrated with the short
horizontal black lines in the dynamic power spectra plots such as
Fig.~\ref{smap_lmap}, centred around the respective strongest
signal.\\
% If the strongest signal is not significant
%($<2\sigma$), we still show the formal error bar
%in period with a dotted short line.\\

We determined the widths of the peaks in the power spectra from
a Gaussian fit to the peak in the power spectrum in frequency
space to compare them to the frequency resolution imposed
by windowing, determined by the duration of the time
segments: d$f=1/T$. The corresponding width in period space
depends on the period $p$, yielding $p^2/T$, thus a narrower
peak for shorter periods. If the measured width of a
significant peak at period $p$ is broader than $p^2/T$, then the
period is not constant.\\

Finally, we computed the fraction of the total time that a
period is detected above a given detection threshold, similar
to a duty cycle, where we determined these fractions for a range
of thresholds. For each source, we computed new time series
of power spectra that do not overlap, and in this sample, we
summed the integration time of all time segments that contain
a signal above a given threshold. We computed the fractions for
each observation separately and for all time segments from all
observations combined. The results are described in
Sect.~\ref{sect:dcycle}.\\

\section{Results}
\label{sect:results}

We performed the analysis described in Sect.~\ref{sect:analysis}
for the observations listed in Table~\ref{tab:obs} and discuss the
results for some of the individual systems in the following subsections.

\subsection{RS\,Oph}
\label{sect:oph}

We searched for the 35s period found from \swift\ observations
by \cite{atel770} in three \xmm\ and two \chandra\ observations of
RS\,Oph. We first analysed the \xmm\ observation taken on day 14
(ObsID 0410180101), which is not listed in Table~\ref{tab:obs}
because it was not taken during the SSS phase; \citep{rsophshock}.
We divided the 23ks observation into 46 overlapping 1000s time
segments but found no significant detection. While a 35s signal
with a false-alarm probability of 10\% is found in one time interval,
the correction for multiple testing reduces the detection
probability to $<1$\%. Thus, repeating a search 46 times in
a light curve without a signal could produce such a peak
and it may thus be a random occurrence. The \xmm\ observation
taken on day 26.1 is also dominated by the early shock emission,
but has an additional soft component that appeared
during the observation. \cite{nelson07} reported that a 35s period
was present during a short time when the soft count rate rapidly
increased by a factor 3, and in Fig.~\ref{smap_lmap}, we give a
graphical illustration of the spectral (left) and period (right)
evolution on the same vertical time axis. Numerous short-lived
unidentified soft spectral emission features, noted by
\cite{ness_obsc}, are included with the values of wavelengths of
unidentified features in the top left panel.
% that need to be
%checked against atomic databases to search for potential
%identifications.
So far, no convincing identifications
were found after investigating possibilities of extreme line shifts
\citep{nelson07} or exotic elemental abundances
\citep{orioposter}.
% (Orio et al.
%2010\footnote{poster presentation \#2.03, during the meeting
%"High-resolution X-ray spectroscopy: Past, Present, and
%Future, March 15--17 2010 in
%Utrecht, The Netherlands (www.sron.nl/xray2010)}).\\

In two time segments around the same time, a periodic signal appeared
at 35\,s above 90\% detection probability, which can be seen in the
right panel. The black line in the top panel is the power spectrum from
the full light-curve and contains no significant signal at 35s.
While this might seem like a marginal result, we emphasise that the
confidence levels and colour scheme have been conservatively corrected
for the effects of oversampling and multiple testing with 24 trials.
Moreover, a similar period value was
obtained with high confidence from later \swift\ and \xmm\
observations, and it also coincides with the rise in brightness
and the appearance of an apparent emission line spectrum with many
unknown line features.\\
%TBD: Possibility that red noise from the rise has caused this detection?\\

Since after day $\sim 29$, the 35s period was always associated with
SSS emission, this coincidence may indicate that the strange soft
component with emission lines (Fig.~\ref{smap_lmap}) is also associated
with SSS emission, even though the spectrum does not appear to be a
typical SSS continuum spectrum. In this context, it is noteworthy that
the low-resolution simultaneous EPIC spectrum can be reproduced by
a blackbody \citep{ness_barcelona}.\\
% (Ness
%2013\footnote{online presentation given during meeting
%"Spanish X-ray Astronomy 2013", Barcelona, June 17--19 2013:\\
%{\tt http://www.ice.csic.es/personal/rea/sxa/2013/Entries/
%2013/6/17\_Final\_program\_files/ness.pdf}}).\\

A \chandra\ observation taken on day 39.7 after outburst
(top panel of Fig.~\ref{rsoph_chan}) contains a peak at 37.5\,s
that proved to be insignificant.
This observation was taken during an episode of low count rate during
an early variability phase \citep{osborne11}.\\

\begin{figure*}[!ht]
\resizebox{\hsize}{!}{\includegraphics{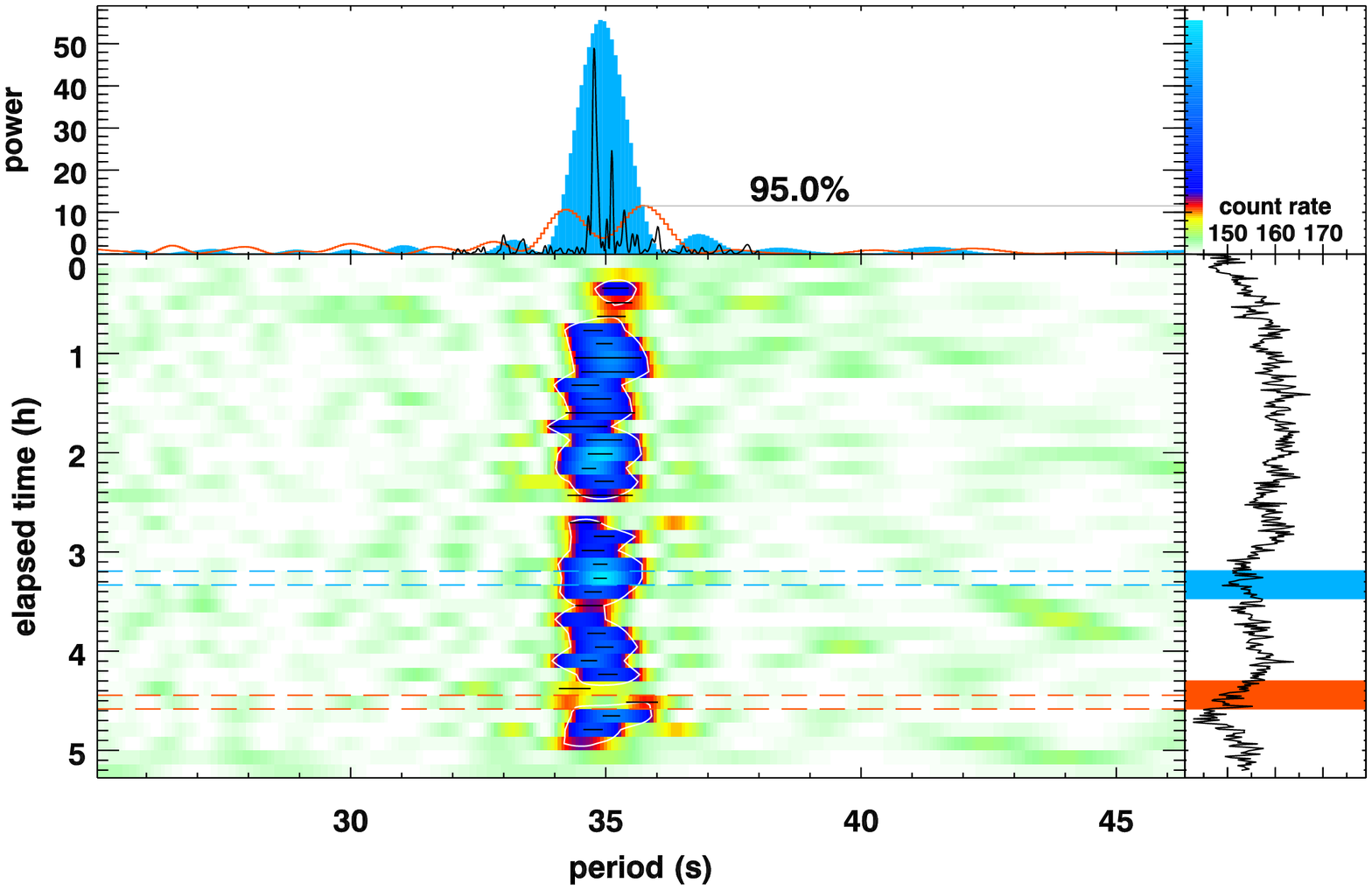}\includegraphics{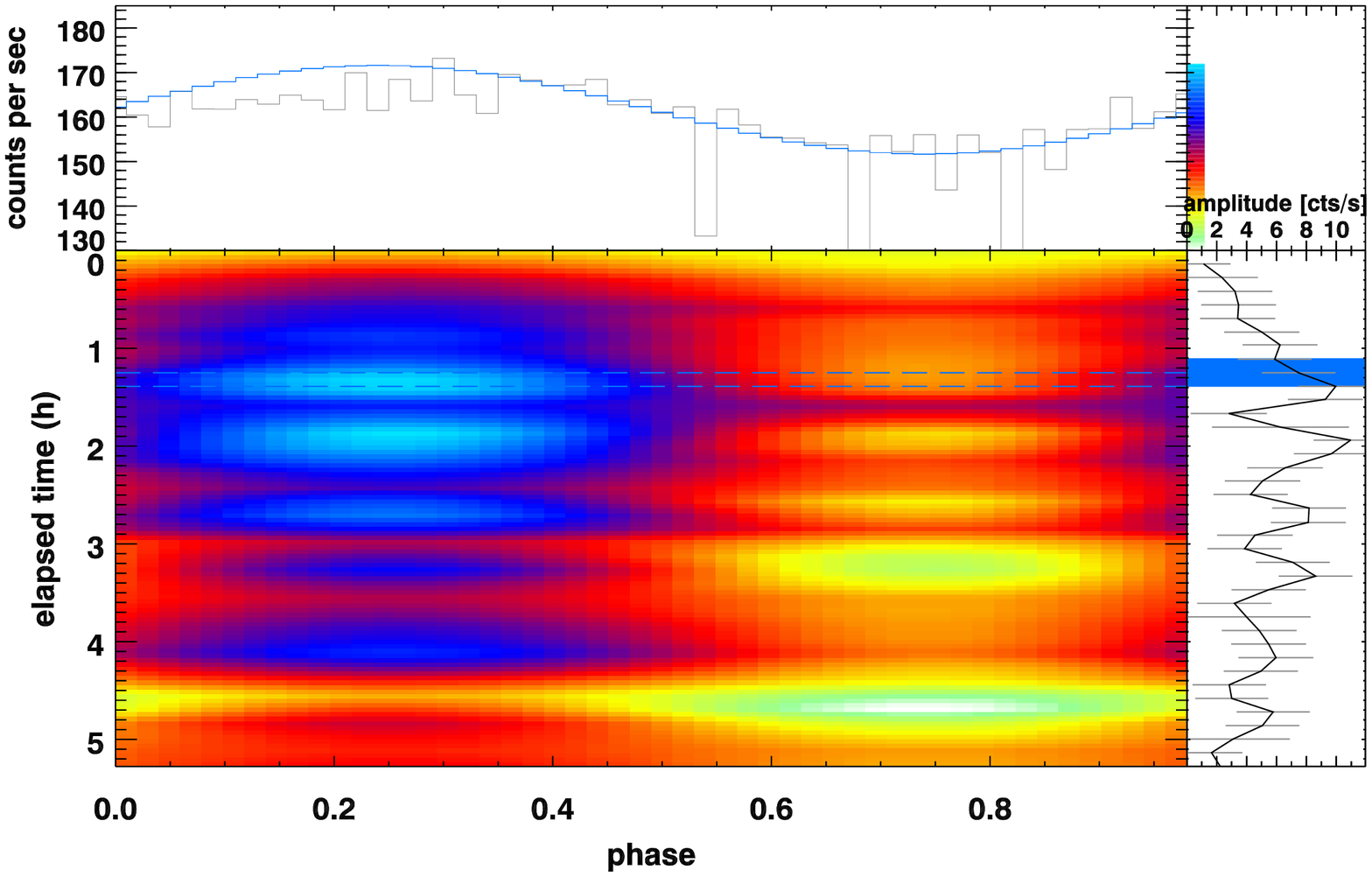}}
\caption{\label{lmap_oph}RS\,Oph on day 54.
{\bf Left}: Time map of the 35s period in the \xmm\ RGS2 light curve,
see description of right panel in Fig.~\ref{smap_lmap}.
% The light curve (shown in the right-hand
%side panel) contains dips that originate from telemetry drop outs that
%occurred in all instruments owing to the extreme brightness. They
%were removed for the period searches.
%The shaded areas in the right panel mark
%the 1000s time intervals in the light curve used to construct the
%power spectra of the same colour used in the top panel, which also contains
%the power spectrum derived from the whole data set (shown by the black
%line). While overlapping
%1000s time segments were analysed, the corresponding power spectra are
%mapped into adjacent 500s intervals in the central panel, thus not
%overlapping.
%The 500s gap between the blue and orange pairs of dashed
%lines corresponds to the power spectra extracted from the overlapping
%time interval (not shown in top). The short black lines in the main
%panel indicate uncertainties of periods (see caption of
%Fig.~\ref{smap_lmap}).
{\bf Right}: Time map of folded light curves, where each curve was
artificially moved in phase for the peak to occur at phase 0.25.
In the top panel,
an example is shown with the sine curve fit from which
the modulation amplitudes with uncertainties were determined.
In the right panel, the evolution of amplitudes is shown.
}
\end{figure*}

During the \xmm\ observation taken on day 54, the peak of the overall
X-ray brightness evolution \citep{osborne11}, the 35s period
was detected at a significance $>99.9$\%.
The extreme brightness of RS\,Oph at this time led to difficulties
in telemetering the data to the ground, and the EPIC data
can only be used for spectroscopy. The RGS1 light curve could not
be used for the same reason; only the RGS2, which had a lower
count rate owing to a failure of a CCD chip (that would have recorded
the dispersed photons from the peak of the SSS spectrum), contained a
controllable number of telemetry drop outs. In Fig.~\ref{lmap_oph} we
show the time evolution of the period (left) and of the amplitude (right)
from the RGS2 data. The signal power varies considerably, as shown by the
two power spectra in the top panel, extracted from the time intervals
marked in the right panel of the left plot. The orange open histogram
contains two much weaker signals at $\sim 34$\,s and $\sim 36$\,s.
In the right panel, the evolution of the phase-folded light curves
is shown on the same (vertical) time axis as the left plot. Since the
start and stop of the time segments do not coincide with the same epochs,
we have artificially moved each curve such that the peak is always
at phase 0.25. The right plot in the right panel shows the evolution
of the amplitude.\\
% while at the same time another weak signal around 44\,s might be present.\\

\begin{figure*}[!ht]
\resizebox{\hsize}{!}{\includegraphics{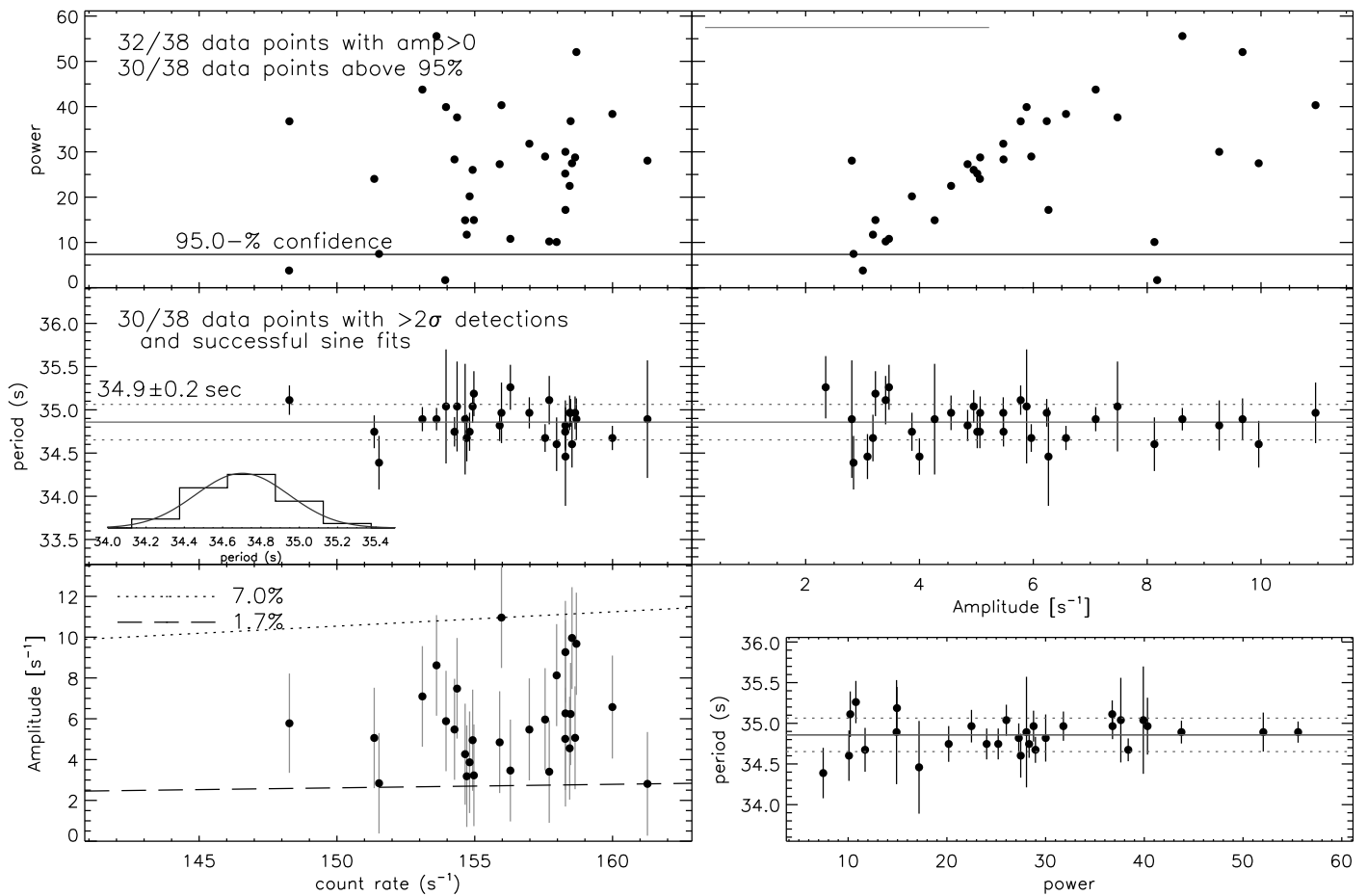}}
\caption{\label{per_cr_oph}RS\,Oph on day 54: Relations between
power (top row) and period (middle row) versus mean RGS2
count rate (left column) and modulation amplitude (right column).
The amplitude is mean to peak in units of RGS2 counts per second. A
typical uncertainty range in amplitude, derived while fitting
sine curves to the phased light curves, is shown with the grey
horizontal line in the left corner.
The 95\% ($2\sigma$) detection level (corrected for oversampling)
is marked by the horizontal line in the top panels. In the {\bf
middle panels}, the error-weighted median period with standard
deviation of the periods is marked with the horizontal
solid and dotted lines, respectively. The 95\% error bars on
periods were calculated with Eq. (4) in \cite{larsson96}
(see text). We show 30 out of 38 data points for which the
signal power exceeds a 95\% detection and a sine fit yields a
positive amplitude. The distribution of periods is shown
in the small inset where a Gaussian is added for comparison.
In the {\bf middle right panel}, period versus amplitudes are shown,
suggesting that larger variations in period occur with smaller
amplitudes. In the {\bf bottom left panel}, amplitudes are shown
versus count rate, demonstrating that there is no correlation.
Ranges of amplitudes relative to count rate are marked by the
dashed and dotted lines. In the {\bf bottom right panel}, periods
versus power are shown, which also seem to show that larger variations
in period occur at lower values of power.
}
\end{figure*}

\begin{figure}
 \resizebox{\hsize}{!}{\includegraphics{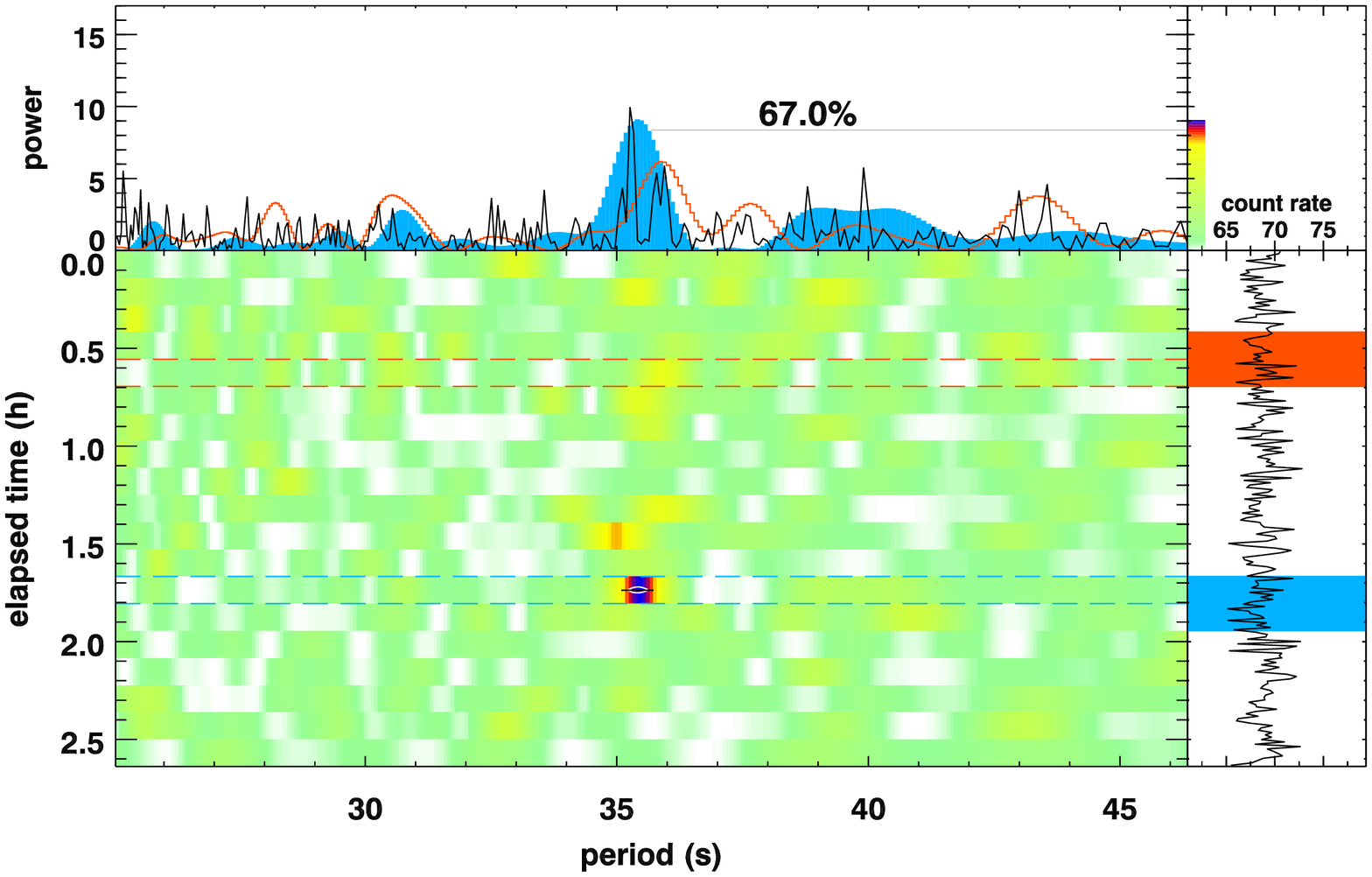}}
\caption{\label{rsoph_chan}RS\,Oph on day 67: Time map around the
35s period in the \chandra\ HRC light curve, see description in
the right panel of Fig.~\ref{smap_lmap}.
The strongest peak coincides with the 35s period seen in earlier
\swift\ and \xmm\ observations (see Fig.~\ref{lmap_oph}). The
power (conservatively corrected for multiple testing and oversampling,
see Appendix) only yields a 1-$\sigma$ detection.
}
\end{figure}

We studied the relations between the period, power, amplitude,
and the mean count rate in 38 time segments of the
full light-curve, excluding six segments in which the amplitude only
yields an upper limit. In Fig.~\ref{per_cr_oph} the values of these parameters
are plotted against each other. In the left column, the three period
parameters are plotted against the mean count rate.
In the top row, the 95\% (2$\sigma$) detection threshold (corrected for
oversampling) is indicated by the horizontal line. The signal power
does not seem to depend on the count rate.
% A Kendall correlation coefficient
%yields a value of 0.282, indicating a weak positive correlation,
%but with a too low significance for any conclusions.
Out of the 32 time segments with positive amplitudes, 30 yield
$>2\sigma$ detections of the 35s period, and these are used in the
two panels below. The 95\%
error bars included for the periods in the middle left panel are
twice the standard deviation, calculated with Eq. (4) in
\cite{larsson96}. The error-weighted mean period is given with
the horizontal line with the value and standard deviation
given above this line. The errors are smaller than the widths
of the peaks, which are driven by windowing and thus the duration
of the time segments $T$. The frequency resolution scales with
$1/T^2$, and at the period of 35s, the expected width of the peak
is d$p=p^2/T\sim 1$s. The variations of the period are slightly
larger than the 95\% error bars and thus may be real, although we
note that the large uncertainties in amplitudes have not been
accounted for when calculating the period errors. Furthermore,
the distribution of periods, shown in the small inset,
indicates that the variations are normally distributed, because the
distribution can be fit by a Gaussian. The distribution and formal
errors do not provide information to conclude whether or not the
period varies with time, as concluded by \cite{cal83_67sec} for
the 67\,s period in Cal\,83. In the top panel of Fig.~\ref{lmap_oph},
the power spectrum derived from the complete light curve is shown
by the thin black line, and the higher period
resolution achieved by the longer baseline of the total light
curve reveals multiple peaks. This would support that the
period is not constant in time.\\

In the bottom left panel of  Fig.~\ref{per_cr_oph}, the amplitudes
in units of RGS2 count rate are shown against the average count rates for
each time segment. The dashed and dotted lines indicate the lowest and
highest amplitudes relative to the respective count rate in per cent. The
amplitudes range between 2.3--11 counts per second (1.7--7\% of the total
count rate) without any systematic trend with count rate. Uncertainties
in amplitude range between 30\% and 90\%.\\
%No modulation amplitude below
%2\% is detected with $>2\sigma$ significance, most likely because
%any smaller amplitudes will become undetectable within the noise.\\

In the right column of Fig.~\ref{per_cr_oph}, relations between
the amplitude, power, and period can be studied. Obviously,
small amplitudes are more difficult to detect because they yield low
detection powers. Some outliers with low signal power but high
amplitude can be caused by a high degree of additional non-periodic
variability, which in a folded light curve could result in a large
amplitude. The uncertainty in amplitude is marked with the
horizonal grey line in the top left panel and is larger than the
deviations from the main trend.
 In the middle right panel, the variations of period
with amplitude are shown and it seems that the larger variations
in period coincide with low amplitudes. This trend seems even stronger
in the bottom right panel, which plots power versus period.
\cite{osborne11} showed that the period stability increased with
time through the SSS phase.\\

Another \chandra\ observation taken on day 67 after
outburst may present some evidence for the 35s period
at the 1-$\sigma$ level (see Fig.~\ref{rsoph_chan}).
The colour scheme and thresholds for testing include the
correction for oversampling and for multiple testing, although
the latter is conservative and does not recognise that this
particular period was clearly detected in other observations.  
\cite{osborne11} reported that the 35s period
disappeared on day 58.8, about a week before this
observation was taken. If the period is real, then
it might be related to a short re-brightening that seems
evident from the \swift\ light curve shown in Fig.~2 in
\cite{osborne11}. On the other hand, the \chandra\ observation
contains no re-brightening, in particular not around the time
of this marginal detection.\\

Only one 4400-sec exposure was taken with the \xmm\ optical
monitor on day 54 with the UVW2 filter. While the exposure was taken
in fast mode, the source was not centred within the small readout
window, leaving most of the PSF outside of it. The light extracted
from the remaining part of the source contains no significant periodic
signal. 

\subsection{KT\,Eri}
\label{sect:eri}

KT\,Eri was the second nova in which a transient 35s
period was seen in \swift\ monitoring observations during the
SSS phase \citep{kteri35}. Four \chandra\ observations were
taken during the eruption on days 71.3, 79.3, 84.6, and 158.8,
and we searched the light curves for periodic oscillations.
Owing to the high brightness of the source and with the main goal
being to obtain high-resolution grating spectra, the observing
times were rather short. In such short exposures, transient
periodicities can be missed easily.\\

\begin{figure}[!ht]
\resizebox{\hsize}{!}{\includegraphics{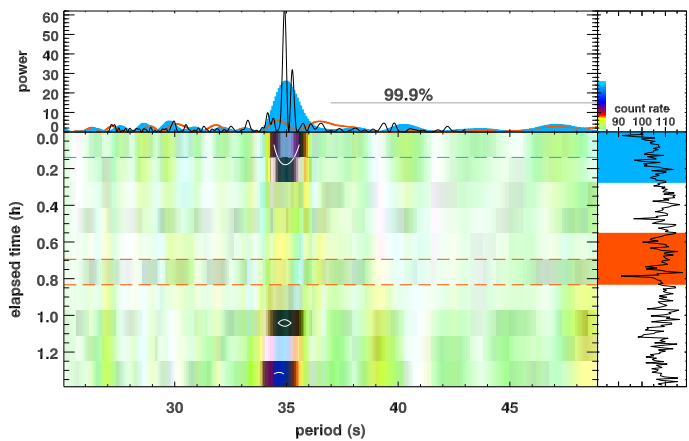}}
\caption{\label{lmap_kteri}KT\,Eri on day 159: Time map around the
35s period in the \chandra\ HRC light curve, see description in
the right panel of Fig.~\ref{smap_lmap}. The thin solid line in the
top panel is the power spectrum from the entire observation and
shows two $>99.9$\% significant peaks. This indicates that the
period is not constant in time.
}
\end{figure}

We detected the 35s period only during the early and late parts of
the observation on day 159 (see Fig.~\ref{lmap_kteri}). 
The width of the peak is consistent with the
highest possible resolution power in period, associated with
the selected duration of each time segment. Extracting the power
spectrum from the total light curve yields a higher resolution in
period, as can be seen from the thin black line in the top panel. 
The double nature of the peak suggests that the period was variable.

%TBD Andy: consistency check with the \swift\ observations
%if there are any simultaneous snapshots.
% is not possible
%because there are no simultaneous \swift\ snap shots with any of
%the \chandra\ observations (Andy: TBD).

\subsection{V339\,Del}
\label{sect:del}

\begin{figure*}[!ht]
\resizebox{\hsize}{!}{\includegraphics{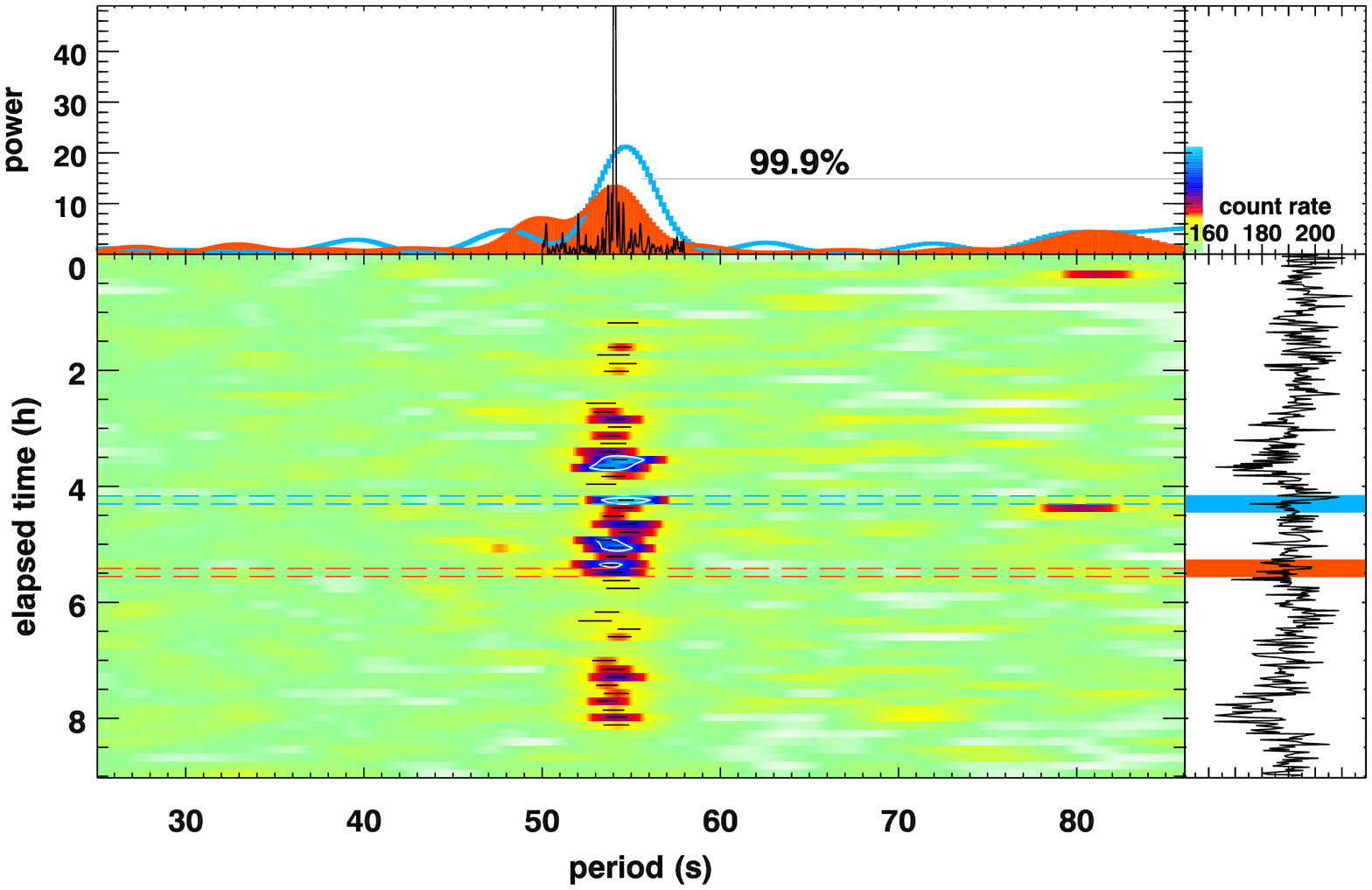}\includegraphics{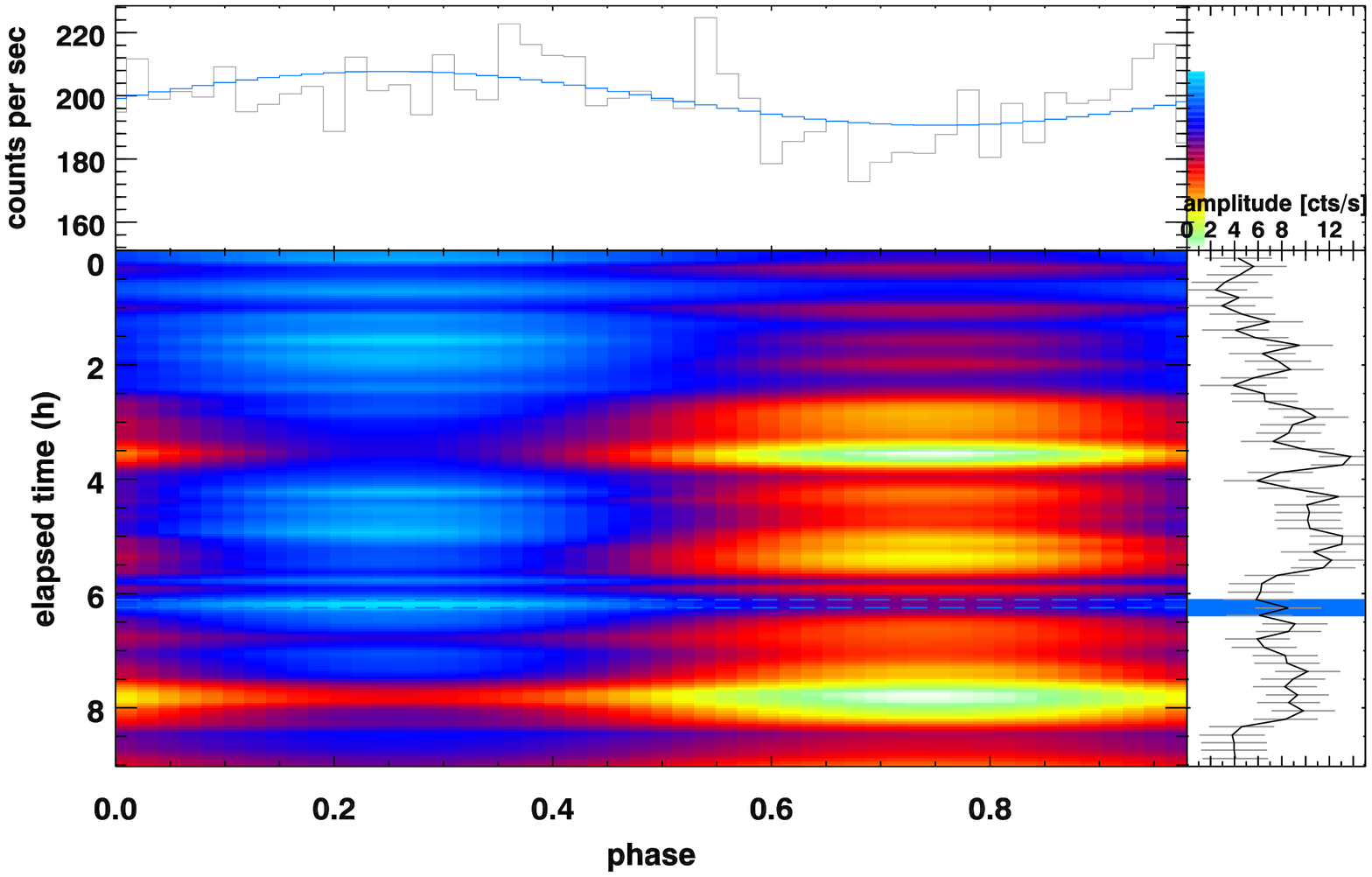}}
\caption{\label{lmap_del}V339\,Del on day 99: Same as Fig.~\ref{lmap_oph},
 focusing
on the 54s period in V339\,Del based on the \xmm\ MOS2 light curve.
The simultaneous combined RGS1 and RGS2 light curve yields a consistent
result.
% The EPIC/pn suffered scientific buffer overlow, while MOS2 only had one drop-out that was removed for the construction of the power spectrum.
}
\end{figure*}

\begin{figure*}[!ht]
\resizebox{\hsize}{!}{\includegraphics{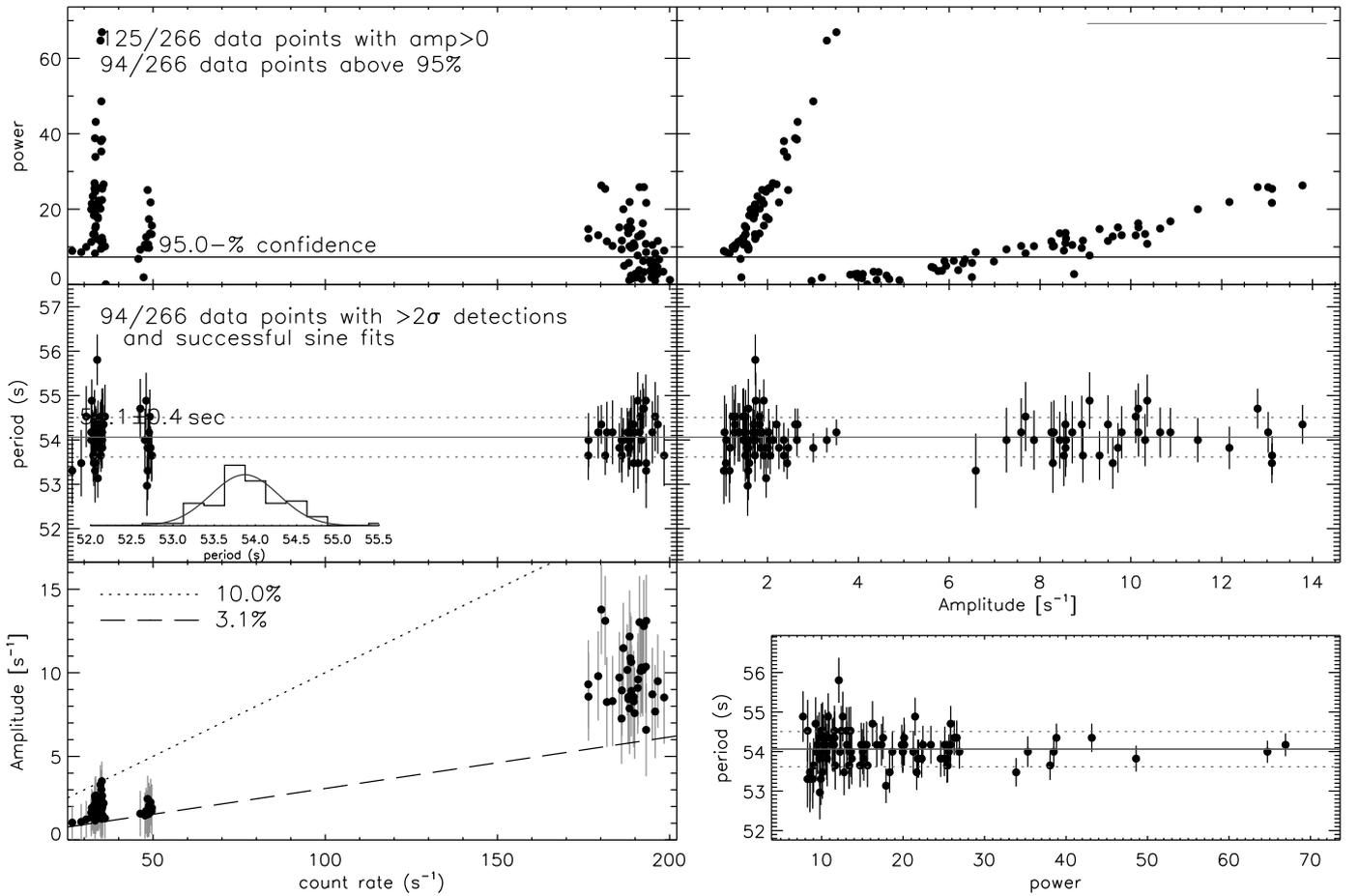}}
\caption{\label{per_cr_del}Same as Fig.~\ref{per_cr_oph}
for the \chandra\ and \xmm/MOS2 observations of V339\,Del.
We show 94 out 266 data points in the middle and
bottom rows that yield $>2\sigma$ detections and positive
amplitudes. As in RS\,Oph
(Fig.~\ref{per_cr_oph}), no relation between period properties
and count rate is apparent, and the period seems to be variable.
The modulation amplitude clearly correlates with the power of the
period because modulations with a larger amplitude are easier to
detect. The \chandra\ and \xmm\ data yield different slopes for
amplitudes in unit counts per second, while for count
rates, a single slope is encountered. In the bottom left panel,
we show that the data from both telescopes yield similar
ranges of relative amplitudes. The bottom right panels appear
similar to RS\,Oph (Fig.~\ref{per_cr_oph}), with largest
variations in period occurring with low values of power.
}
\end{figure*}

The third time a transient short-period signal was seen in
\swift\ observations of a nova was V339\,Del (on days 77.5 to 88.6),
in which \cite{v339del_35} reported a 54s period; this was confirmed
to have also been present in an \xmm\ observation taken on day 99
\citep{ness_v339del}. While the \xmm\ EPIC-pn camera suffered a
full scientific buffer at almost all times and could be used only
for spectroscopy, simultaneous light curves from the MOS and
RGS instruments provide independent evidence. The time evolution
of power spectra derived from the MOS2 data is illustrated in
Fig.~\ref{lmap_del}, where we can see that the signal was not
constant in power, similar to RS\,Oph. We also analysed the two
\chandra\ observations taken on days 87.2 and 114 and found similar
results. We performed the same parameter studies as for RS\,Oph for
all time segments from all three observations and show the results
in Fig.~\ref{per_cr_del}. The \xmm\ and \chandra\ data can clearly
be distinguished by the lower \chandra\ count rates - owing to the
smaller effective areas - and consequently smaller amplitudes, which
are given in units of count rate. The minimum and maximum ratios
of amplitudes versus count rates are indicated in the bottom left
panel with dashed and dotted lines, respectively. The
relative amplitudes are consistent for \chandra\ and \xmm\ data.
In V339\,Del, as in RS\,Oph, no clear
relation between the power and the brightness of the source can be
found. The variations in period
are slightly larger than the 95\% uncertainties in period and
thus  may likewise be real.
The modulation amplitude of signals detected with more than 95\%
confidence varies between 3.1--10\% of the count rate, clearly
yielding more significant detections (higher powers) with
higher amplitudes (top right panel of Fig.~\ref{per_cr_del}).
The slope in the top right panel is steeper for the \chandra\ than
the \xmm\ data owing to the difference in sensitivity and
background.
Again, like RS\,Oph, the amplitude variations are not related to
the brightness of the source (bottom left panel). The bottom right
panel suggests that the largest scatter in period is found
for the lowest power.\\

The \xmm\ optical monitor was operated in fast mode, taking ten
exposures with the UVW2 filter. The count rate ranged around
$140\pm10$\,s$^{-1}$, consistent with Poissonian noise, and we found
no periodic oscillations.

\subsection{LMC\,2009a}
\label{sect:lmc}

\begin{figure*}[!ht]
 \resizebox{\hsize}{!}{\includegraphics{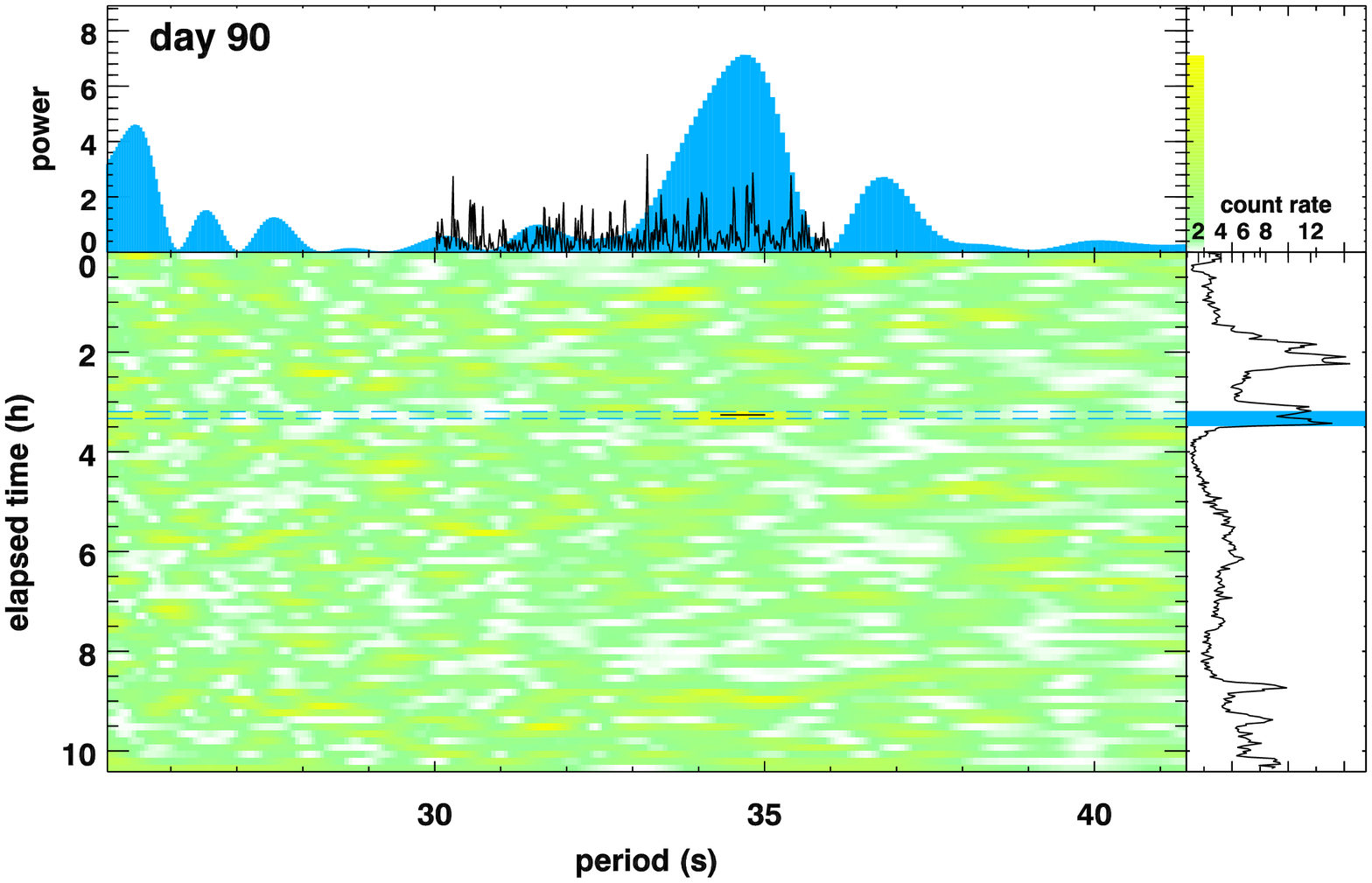}\includegraphics{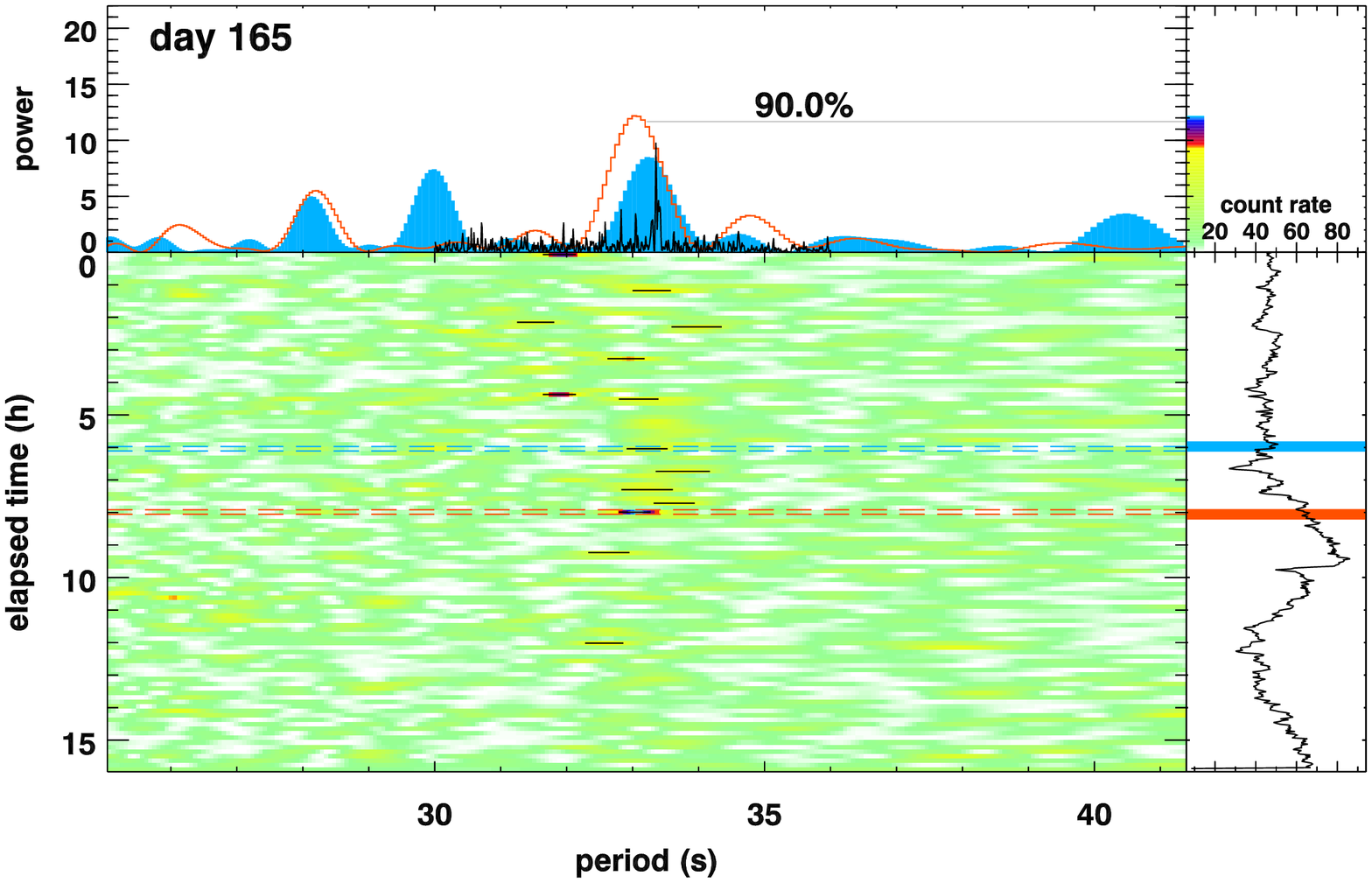}}

 \resizebox{\hsize}{!}{\includegraphics{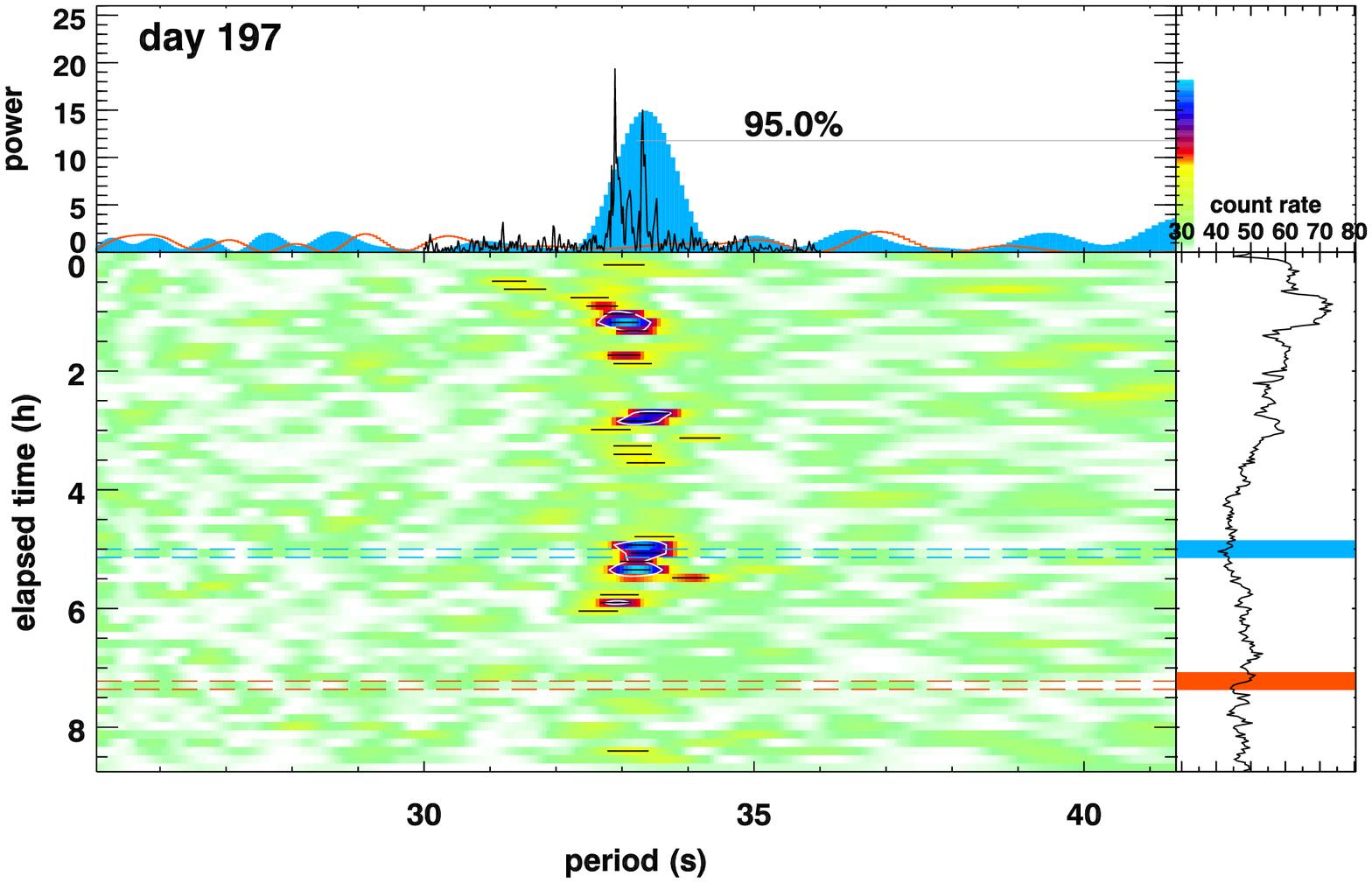}\includegraphics{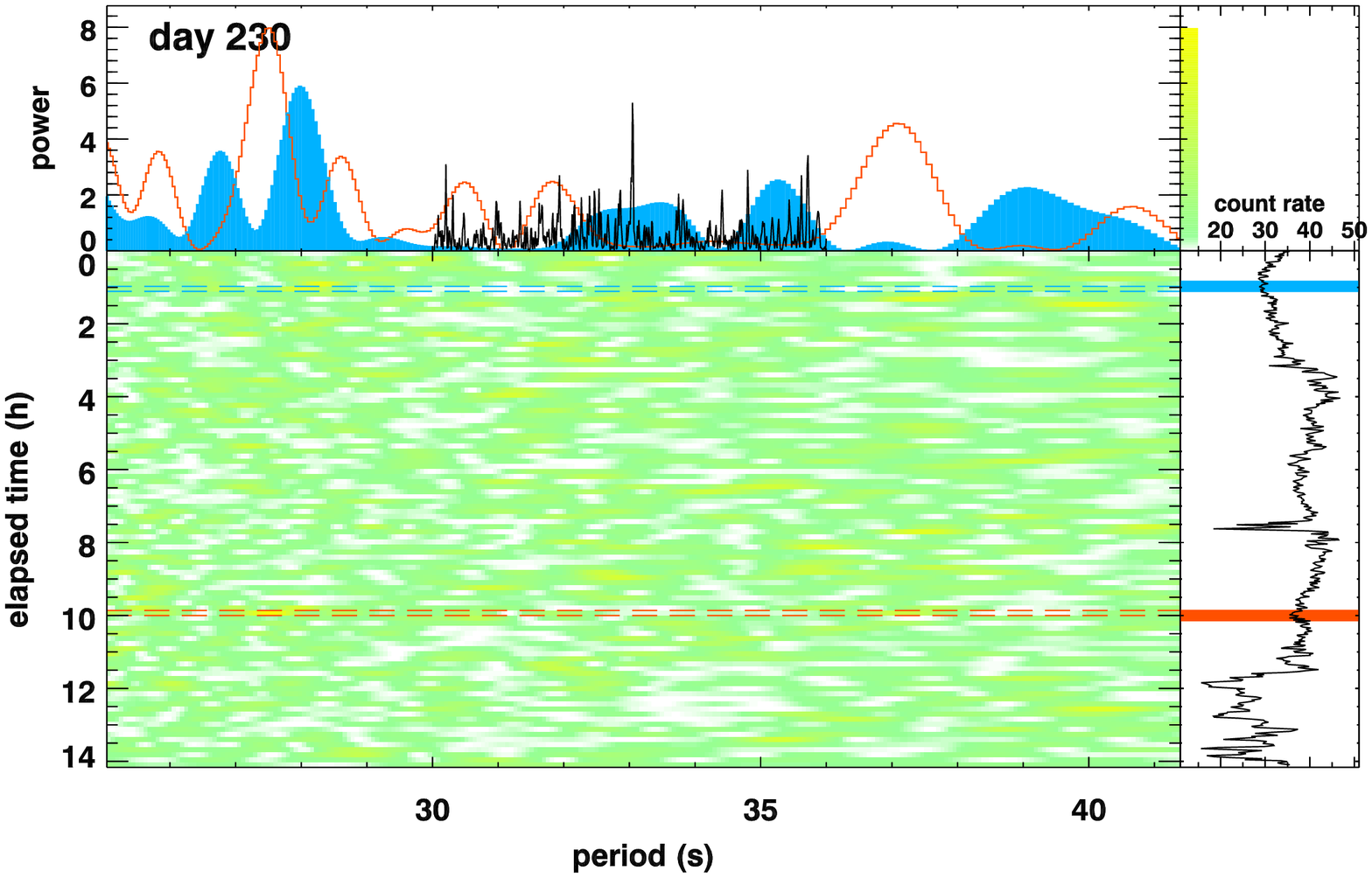}}
\caption{\label{lmap_lmc}LMC\,2009a on days 90, 165, 197, and 230:
Time maps around the 33s period in four \xmm/pn light curves, see
description in the right panel of Fig.~\ref{smap_lmap}.
The vertical (time) axes are different in each plot; they only cover the
respective exposure times.
}
\end{figure*}

%During our systematic search through all observations of bright novae
%during their SSS phase, w
We found a new system with a 33s period in two out of four
\xmm\ observations of nova LMC\,2009a \citep{ATel6147}.
In Fig.~\ref{lmap_lmc}
the time maps are shown for all four observations. In the first
observation, taken on day 90 after outburst (top left panel), no periodic
signal is seen. In the second observation, taken on day 165 after
outburst (top right panel), a significant signal with a period of 33s can
be seen at $\sim 8$ hours of elapsed time.
The same period is more clearly detected in the third
observation taken on day 197 (bottom left panel). In the last observation,
none of the time segments contains a significant signal. We note the
single peak in the power spectra derived from the respective full
light-curves. While a peak at 33s is apparent, it is not statistically
significant. In the bottom
left panel, the total power spectrum contains multiple peaks around
33s, clearly indicating strong variability of the period.
We performed the same studies on correlations as those shown in
Figs.~\ref{per_cr_oph} and \ref{per_cr_del} using 22 data points with
$>2\sigma$ detections and positive amplitudes and find similar results
of significant variability in the period, while variations in power
and amplitude do not correlate with the count rate.\\

As can be seen from Fig.~\ref{lmap_lmc}, the fraction of time that the
33s period is detectable in LMC\,2009a is much lower than for the periods
in RS\,Oph, KT\,Eri, and V339\,Del (see Sect.~\ref{sect:dcycle}).
Perhaps this explains why it is not found in the \swift/XRT data
(Beardmore in prep); \swift\ might have observed only during the
times when the period was absent.\\

Each observation was accompanied by 20-30 fast mode exposures with the
optical monitor, all using the UVW1 filter. The count rates decreased with
time starting from $6.3\pm1.3$\,s$^{-1}$ on day 90.4 through
$2.4\pm0.7$\,s$^{-1}$ on days 165 and 196 to $2.1\pm0.6$\,s$^{-1}$ on
day 230. From each light curve we constructed the power spectrum, but
found no significant signal, even when combining all exposures into
a single light-curve.

\subsection{Cal\,83}
\label{sect:cal83}

\begin{figure}[!ht]
 \resizebox{\hsize}{!}{\includegraphics{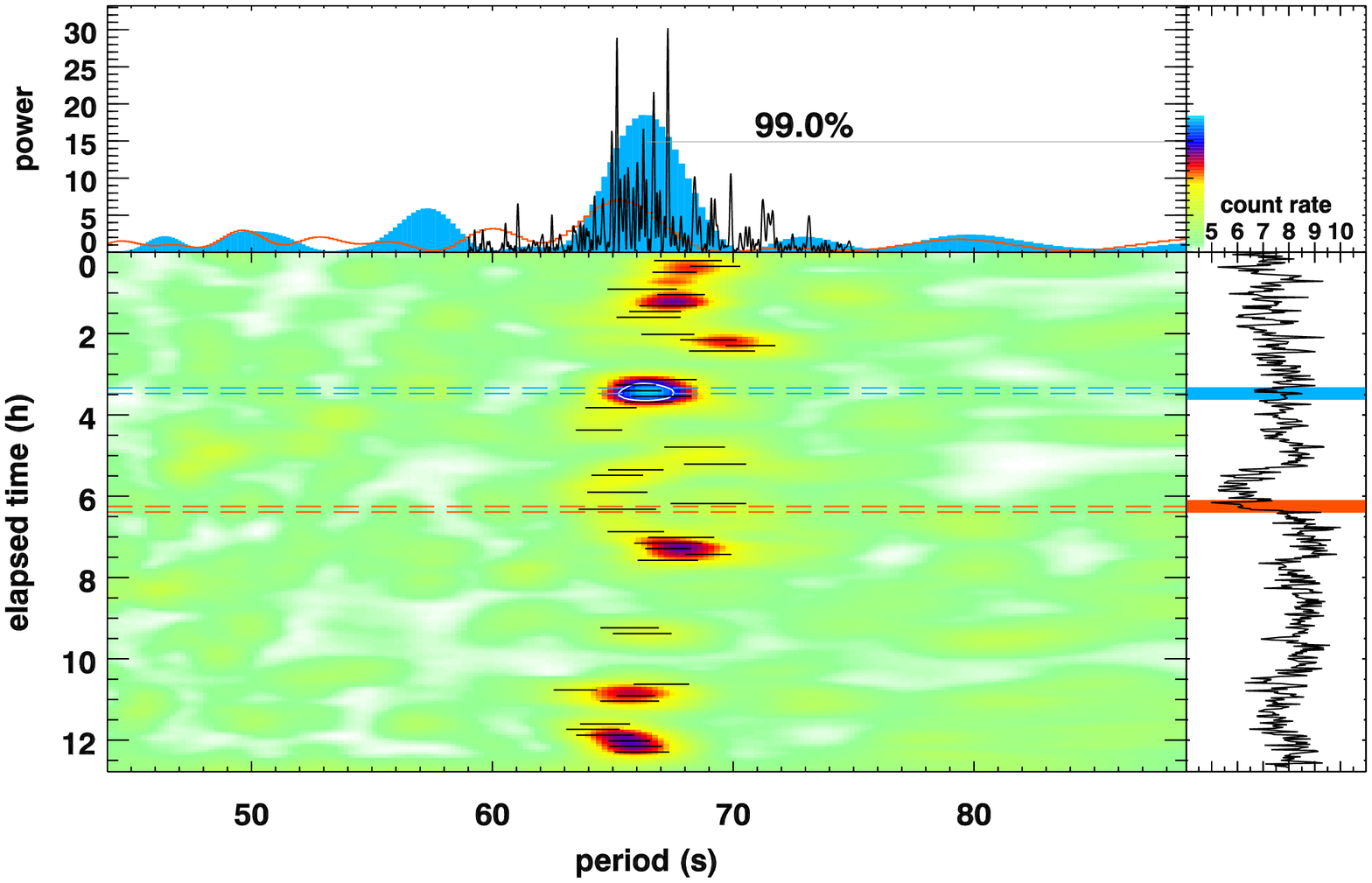}}
%
% \resizebox{\hsize}{!}{\includegraphics{cal83_cr.eps}}
\caption{\label{lmap_cal83}Cal\,83 on 2009-05-30.4: 
Time map around the 67s period in one of 19 \xmm/pn light curves, see
description in the right panel of Fig.~\ref{smap_lmap}.
The time evolution is consistent with that
shown in a different format in Fig.~8 in \cite{cal83_67sec}.
%{\bf Bottom}: Condensed time evolution of count rates in all 19 \xmm\
%observations of Cal\,83 used in this work (see Table~\ref{tab:obs}).
%The horizontal axis represents a time bin number, removing the
%times between observations. Continuous light curves from individual
%observations can be identified by the alternating light/dark grey
%shaded areas. Note that different optical blocking filters have been
%used, for example, lower count rates can be related to use of the thick
%filter (see 4th column of Table~\ref{tab:obs}). The blue circles
%indicate the presence of the 67s period, where the symbol sizes
%scale with the power of the signal.
}
\end{figure}

For the persistent SSS Cal\,83, \cite{cal83_67sec} have recently reported
a 67s periodic modulation. We have produced our period time maps for 19
\xmm\ observations, for which \cite{cal83_67sec} report that the source
was not in the off state, and present an example in
Fig.~\ref{lmap_cal83} for ObsID 0506531701 that yields the same result as
that shown in Fig.~8 of \cite{cal83_67sec}.
% The time
%series of all \xmm\ observations is shown in the bottom panel, where
%the sizes of the blue circles scale with the signal powers. Note
%that count rates are not corrected for the optical blocking filters
%applied. The filters and modes are given in the 4th column of
%Table~\ref{tab:obs}.\\

\begin{figure*}[!ht]
\resizebox{\hsize}{!}{\includegraphics{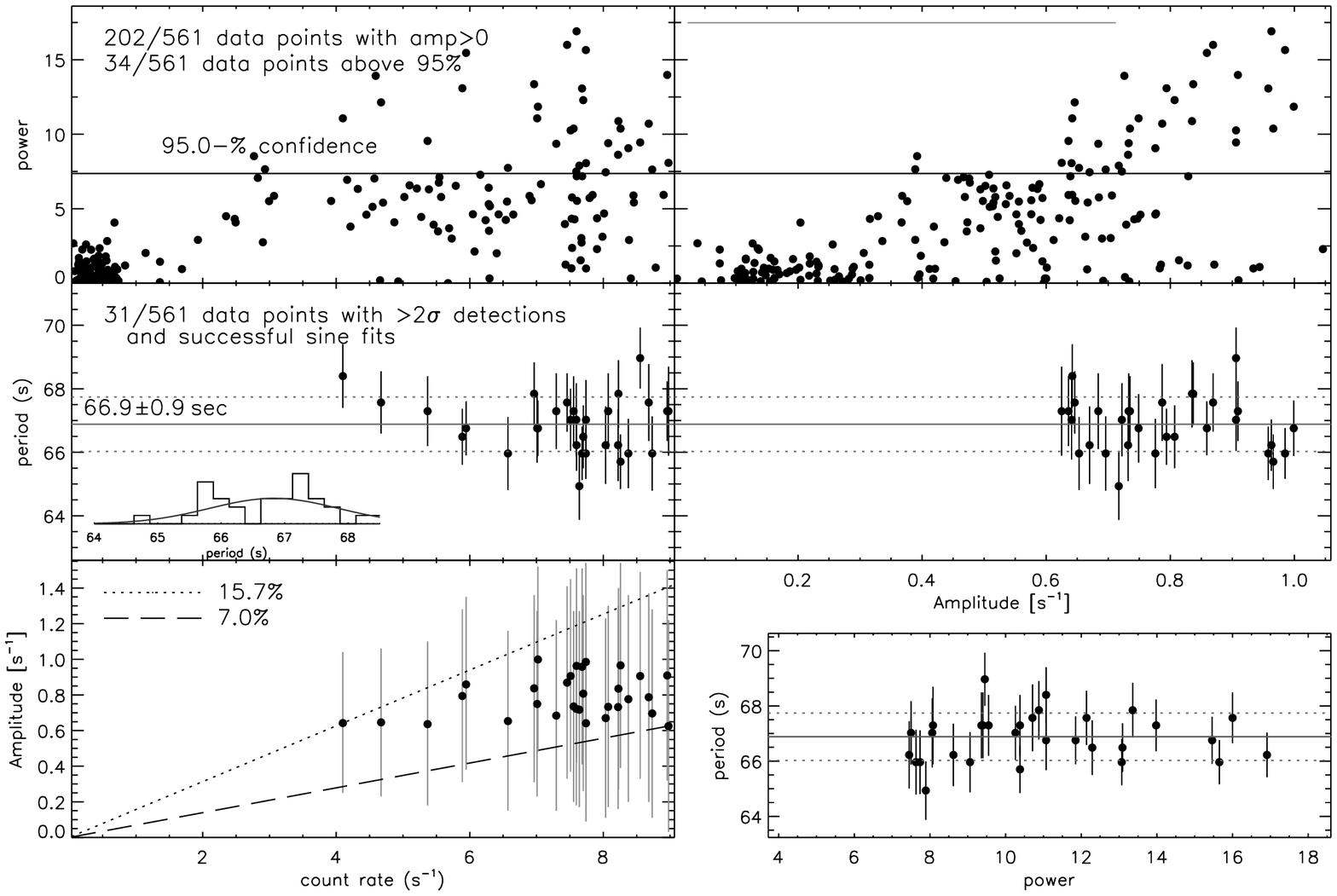}}
\caption{\label{per_cr_cal}Same as Fig.~\ref{per_cr_oph},
with data collected from 20 observations of Cal\,83, divided
into a total of 578 1000s time segments. 17 of these intervals
yield zero count rate and were excluded. Furthermore, all data points
yielding only upper limits in amplitude were excluded, which
left
202 data points, of which 34 time intervals yield a $>95$\%
detection of a periodic signal. The expected correlation between
modulation amplitude and signal power (top right) is contaminated
by a large number of overestimated amplitudes that are most likely
caused by additional, non-periodic variability.
These deviations are consistent with the large uncertainties in
amplitude. The 95\% threshold is corrected for oversampling, but
not for multiple testing. 31 of 34 data points that are found above
this threshold and yield a successful sine fit are used in
the lower two rows. The variations in period of about $\pm 3$s
reported by \cite{cal83_67sec} can be recognised in the middle
row and bottom right, but do not seem to correlate with count
rate, amplitude, or signal power.
}
\end{figure*}

We repeated the same correlation studies of the power of
the period and period with count rate that we have
presented for
RS\,Oph and V339\,Del. For Cal\,83, we have a much larger
sample, which is shown in Fig.~\ref{per_cr_cal}. In the top
panel, we include the results from all time segments.
%, but excluding those times with zero count rates caused by gaps in the data.
The variations in period are again slightly larger than
the 95\% errors and thus might be real. At intermediate
count rates, the period seems to be anti-correlated, but
this would be based on too few data points (middle row). We
also searched for periodic variations of the
power on amplitude on longer time scales, but found no
clear patterns in the power spectra of signal powers and
amplitudes.\\

We checked all available \xmm\ observations with the optical
monitor, which contain exposures in U, B, UVW1, UVM2, and UVW2 filters,
but found no periodic signal in any of these exposures nor in
combined light curves for each filter.

\subsection{Search for other candidates}
\label{sect:other}
 
%In the appendix section (online version?), we include the time maps
%for all \chandra\ and \xmm\ observations in our
%sample listed in Table~\ref{tab:obs} that are not discussed
%in the previous sections.

We screened all power spectra from the datasets listed in
Table~\ref{tab:obs} for any significant signals
that  might account for multiple testing and oversampling. With these
corrections any peak at any period that rises above a given
threshold is a strong candidate. We searched for signals
that rise above the 95\% threshold, but found only the five
sources already discussed above, thus no other candidate 
can be identified.\\

\subsection{Fraction of observing time for which short-period
oscillations are detectable}
\label{sect:dcycle}

To compute the fraction of time for which the signals 
are detectable above a given threshold, we recomputed power
spectra for adjacent, non-overlapping time segments of 1000s
duration. We first defined a detection threshold and summed
the exposure times of all time segments in which a period
is detected above the given threshold. We then computed the
ratio of the summed exposure time by the total available
exposure time. The resulting fractions depend on the chosen
significance level, yielding higher fractions for lower
detection thresholds.\\

\begin{figure}[!ht]
\resizebox{\hsize}{!}{\includegraphics{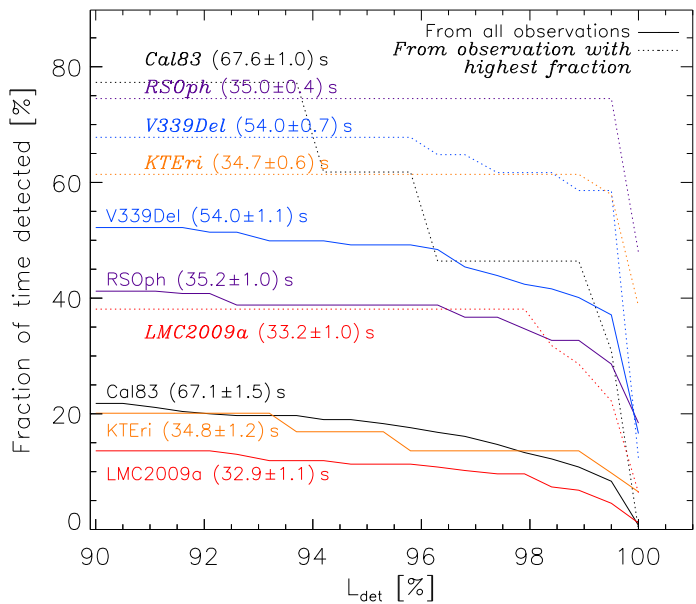}}
\caption{\label{dcycles}Fractions of observing time for which the
strongest periods are detectable as a function of detection threshold.
The solid lines are the average fractions of time derived from
all available observations, while the dotted lines (with labels
in italics) are the fractions of time derived from the single
observation in which the period is seen most frequently.
For this plot, power spectra from directly adjacent,
non-overlapping time intervals have been used. 
For no system, the period was detected 100\% of the time,
showing the transient nature. Furthermore, the signal is episodic
because the fractions are not the same in all observations.
}
\end{figure}

For each target, we probed a range of detection levels, and in
Fig.~\ref{dcycles} we show fractions of times as a function of
detection threshold for the five sources with detected periodic
signals.
Since the period was not necessarily present in all observations,
the global ratio strongly depends on the coverage of the total
evolution, which is generally much better determined with \swift.
To study the short-term behaviour, we also include
in Fig.~\ref{dcycles} the ratio for the observation with the
highest fraction with dotted lines and labels in italics for each
source. For all sources, solid lines indicate that the transient
signal is episodic. These estimates can be refined with the
\swift\ data, which will be presented by Beardmore et al. (in preparation).\\

\subsection{Search for patterns in the transient signal}
\label{perper}

The short-period oscillations are clearly transient in all five
systems with detected periods, and we investigated whether there
are any patterns in the power of the signal. For example, periodic
variations could be caused by the rotation of the white dwarf
if the oscillations originated in an accretion spot. For each
source, we created a time series of the signal power and
period value from all time segments.
%An example of
%such a time series for Cal\,83 can be seen in Fig.~\ref{lmap_cal83},
%where the symbol sizes scale with the power of the 67s signal.
We computed a power spectrum based on the central
times of each time segment and the associated parameters.
None of the resulting power spectra reveal any significant signatures.

\section{Discussion}
\label{sect:disc}

We have four confirmed novae with periodic or quasi-periodic
transient signals ranging between 33 and 54\,s and one persistent
SSS with a periodic signal of 67\,s. For these five systems, we list in
Table~\ref{tab:withper} the periods, limiting count rates, and
relative amplitudes of the modulation for 2-$\sigma$ detections,
the optical speed class parameters $t_2$ and $t_3$, and fractions
of time when detected. The limiting count rates and amplitudes are
the lowest value from the sample of time intervals when a period was
detected with more than 2$\sigma$ significance. Although this
sample is too small for searches of systematic trends, it covers
a wider range of parameters than the time series of individual
systems illustrated in Figs.~\ref{per_cr_oph}, \ref{per_cr_del},
and \ref{per_cr_cal}. No systematic trends can be seen between the
period and the count rate or amplitude, but the nova speed class seems
to scale with the period, yielding longer periods for slower
novae. The speed class is known to scale with the mass of the underlying
white dwarf, yielding higher masses for shorter speed class
\citep{livio92}, and shorter periods thus may be present in
systems with more massive white dwarfs.
%The masses in the last column have been derived from $t_3$ using
%Eq. (13) from \cite{livio92}.\\

\begin{table*}
\begin{flushleft}
\renewcommand{\arraystretch}{1.1}
\caption{\label{tab:withper}Properties of systems with known periodic signals}
{%\scriptsize
\begin{tabular}{llllllll}
\hline
  Target &
  period$^{a}$ &
  ${cr}_{\rm lim}^{b}$ &
  A$^{c}$ & 
  A$^{c}$ &
  $t_2$ & $t_3$ &
  duty cycle$^d$ \\
  & (s) & (s$^{-1}$) &  (s$^{-1}$) &(\%) & (days) & (days) &(\%)\\
LMC\,2009a & $33.2\pm0.1$  & 45 & 1.5--2.9 & 2.9--4.8 & 4$^{(1)}$ & 11$^{(4)}$ & 11.3/38.1\\
RS\,Oph & $34.9\pm0.2$ & 148& 2.3--11 & 1.5--7.0 & 7.9$^{(2)}$/6.2$^{(5)}$ &17.1$^{(5)}$ & 38.8/74.5\\
KT\,Eri & $35.0\pm0.1$ & 70 & 1.5--4.2 & 2.5--3.9 & 6.6$^{(2)}$/6.2$^{(6)}$& 14.3$^{(6)}$ & 16.9/61.4\\
V339\,Del & $54.1\pm0.3$ & 176 & 6.6--13.8 & 3.4--7.6 & $12\,\pm\,2^{(3)}$ & 23.5$^{(3)}$ & 49.2/67.8\\
Cal\,83 & $66.8\pm0.5$ & 4.0 & 0.6--1.0 & 7.0--15.7 & -- & -- & 18.7/61.8\\
\hline
\end{tabular}
}

$^{a}$Errors on periods are the standard deviation.\\
$^{b}$Minimum count rate during time intervals when periods with $>2\sigma$ are found\\
$^{c}$Amplitude range in counts per second (Col. 4) and in \% of count rate (Col. 5) that yields $>2\sigma$ detections\\
$t_2$ and $t_3$: from (1) \cite{schwarz2011}, (2) \cite{hounsell10},
(3) \cite{v339del_t2},
(4) \cite{lmc2009a_t3}, (5) \cite{rsoph_t3}, (6) \cite{kteri_t3}
%$^a$White dwarf mass derived from $t_3$ using Eq. (13) from \cite{livio92}.
\\
$^d$Fraction of observation time (all observations combined/in the single
observation with highest fraction) detected at $>95$\% ($2\sigma$)\\
\renewcommand{\arraystretch}{1}
\end{flushleft}
\end{table*}

%Questions to address:\\
%1) When is the period observable?\\
%2) Under which circumstances is it detectable (count rate)?\\
%3) What makes periods weaker?\\
%4) Based on the 5 systems, make predictions and test with tested
%novae; is anything weaker at the predicted periods?

%\subsection{Lowest required count rate to detect signal}

%The top panel of Fig.~\ref{per_cr_cal} contains all time
%segments of Cal\,83 from which one can see that no
%significant detection occurs when it has faded to a
%count rate below $\sim 3.5$\,cps.

%TBD: Andy, Some theoretical calculations of what type of
%minimum count rate would be needed (assuming Poissonian noise)
%to actually see this period?

\subsection{Detectability of the period during Thomson scattering?}
\label{sect:thomson}

In this section we show that Thomson scattering does not
reduce the detectability of the period.
Recent studies of the eclipsing recurrent nova U\,Sco have
revealed that about 50\% of the continuum emission from the
white dwarf is still visible during total eclipse. The shape
of the continuum spectrum seems to have been preserved, while
emission lines dominate the spectrum. This indicates
substantial effects of Thomson scattering of the continuum
while the emission lines are not eclipsed \citep{ness_usco}.
A few other SSS spectra show similar characteristics of
reduced but detectable continuum emission with emission
lines on top, and
\cite{ness_obsc} have classified this type of SSS
spectrum as the SSe subclass: 'e' for emission lines as
opposed to SSa with continuum-dominated spectra in which
deep absorption lines can be seen.\\
% Some of the observations
%analysed in this paper emit SSe spectra but none of them
%show short-period oscillations. In order to assess whether
%they might be smeared out during Thomson scattering, we
%present the following considerations.

\cite{changkylafis83} have studied the effects of
scattering of photons on the spectra and variability
patterns of X-ray sources. Since Thomson scattering is an
energy-independent process, their calculations of spectral
changes are not of interest to us, and we focus on their
predictions of reductions of the amplitude of oscillations
caused by
scattering. These reductions are shown in their Figs. 7 and
8 and depend on the ratio of the period $p$ to the light
travel-time, $p$\ c/$r_0$ and the optical depth,
$\tau=\int_0^r n_e \sigma_t {\rm d }r$
with $\sigma_t=0.6652457\times 10^{-24}$\,cm$^2$ the cross-section for Thomson scattering by electrons, and $n_e$ the
electron density. For our purposes, we focus on the
models for a uniform sphere because we most likely examine
a continuous outflow and not an abrupt ejection event that
would produce a thin shell. We have observed amplitudes of
about 10\% of the count rate and can thus concentrate on the
range of amplitudes viewed from infinity log$(A_\infty)>-1$
in Fig. 8 in \cite{changkylafis83} and
thus log$(p/t)>0.5$ with $t=r_0/$c the light travel-time.
While this applies only to the case $\tau=10$, lower values
of $\tau$ yield some 0.2 dex lower values of
log$(p/t)$ for log$(A_\infty)>-1,$ which can be seen from
Fig.~7 in \cite{changkylafis83}. We thus focus on the
range log$(p/t)$>0 and thus $r_0 < {\rm c}\ p$.
For the period range 33--67\,s, this constrains the
size of the scattering region to $r_0<($14--29$){\rm R}_\odot$.

%Each scattering event causes a displacement of a photon
%in phase leading to a phase shift of $\delta_l/({\rm c}\ p)$
%where $\delta_l$ is the change in the path length the photon has
%to travel if it scatters. Assuming that we have a white dwarf
%emitting at the centre of a scattering sphere radius $r$,
%$\delta_l$ will depend on $r$ and the number of scattering
%events before escape.
%If we assume about one scattering for each photon and
%ignoring double scattering, a typical $\delta_l$ will then be
%a length of the order of the size of the sphere, but will be
%a random number $0 \leq \delta_l \leq 2r$, likely peaking at
%$\sim r$. The lower limit applies to a very small scattering
%angle while $2r$ is for a photon that travels all the way in
%the opposite direction before being scattered by 180 degrees
%at the far edge of the region.\\
%
%For $r << {\rm c}\ p$, $\delta_l << {\rm c}\ p$, the phase
%shift is $<< 1$ for all scattered photons - i.e., the coherence
%of the period is preserved. If, on the other hand,
%$r \sim {\rm c}\ p$, then the average phase shift is of order
%1. While an exact phase shift of 1 would preserve the coherence,
%the scattering is a random process distributed between 0 and 2,
%and the next phase coherence is degraded, effectively leading
%to a reduction in the amplitude.\\

Depending on the assumed optical depth $\tau$, this allows us
to place a constraint on the electron density $n_e$ of this
plasma. With $\tau = n_e \sigma_t r$ for a uniform sphere
and $r < {\rm c}\ p$, the constraint on density is thus
$n_e>\tau/(\sigma_t\ {\rm c}\ p)$.\\

\begin{figure}[!ht]
\resizebox{\hsize}{!}{\includegraphics{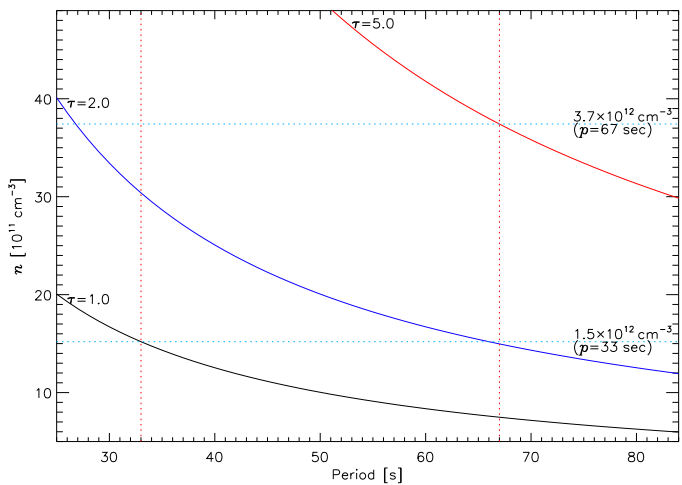}}
\caption{\label{thomson}In the presence of Thomson scattering,
the observable amplitude of a signal will be lower than the original modulation amplitude, depending on optical
depth, $\tau$, and density, $n$. We show estimated densities
below which periodic oscillations would lead to reductions of the
amplitude (and thus detectability) by 64--99\%, depending
on three values of optical depth $\tau=$5--1 that would
lead to 1--37\% of photons escaping unscattered; see
Sect.~\ref{sect:thomson}.
}
\end{figure}

In Fig.~\ref{thomson}, the derived densities for a range
of optical depths $\tau=$1--5 (assuming that 1--37\% of photons
escape unscattered) are illustrated for a period range of 25--85\,s.
In the extreme case of an intrinsic amplitude of 100\%, Thomson
scattering can reduce the amplitude to 10\% of the count rate
only if the electron density of the scattering medium exceeds
$10^{13}$\,cm$^{-3}$. If the intrinsic amplitude is much below
100\%, then the required densities would have to be even higher.
A known intrinsic pulsed fraction and optical depth $\tau$ would
allow us to work out both
the size and density of the scattering region by the signal
contrast degradation, but the intrinsic
amplitude is probably not much higher than the observed one.\\

We thus conclude that Thomson scattering does not effectively
reduce the detectability of the periodic signal, and in fact we
also found a periodic signal in the SSe LMC\,2009a, even though
it is the system with the smallest fraction of total observed
time with significant detections and the smallest amplitude.

\subsection{Rotation period of the white dwarf}
\label{disc:rot}

The long-term presence of the 67s X-ray period in Cal\,83
has led \cite{cal83_67sec} to consider
the possibility that the rotation period of the white dwarf is
the main driver of the oscillations. While typical rotation
periods of white dwarfs are much longer, the fastest
rotation period in a white dwarf binary, RXJ0648.0-4418/HD49798,
was observed to be 13.2s \citep{fastwd}, and a rotation period
of 67s is thus possible. \cite{cal83_67sec}
argued that a long mass accretion history allows the white dwarf to
be sufficiently spun up. The greater challenge would be to explain
the observed changes in period by $\pm 3$s. Because of the large amount
of inertia of a white dwarf, this drift cannot be the result of
accretion-induced spin-up or -down. In addition, there is no relation with
the orbital period of $1.0475\pm0.00004$ days
\citep{schmidtke04,cal83_67sec}, which excludes Doppler shifts
imposed by the orbital motion of the X-ray emitting plasma, for
example. Furthermore,
the possibility was discussed that the observed period is the beat
between the rotation period of the white dwarf and the Keplerian
period of the denser X-ray emitting plasma, but the required rotation
period of the white dwarf would then have to be even shorter, with
an estimated value of 4--12\,s, which would set a new record, although this is still
longer than the break-up period of 3.4\,s. Finally, \cite{cal83_67sec}
argued that the X-ray emitting plasma must originate from an
extended envelope, some two to three times the radius of the white dwarf,
and if that envelope does not rotate synchronously, the spread in the
67\,s oscillations can result. However, this is not quantified,
and the cause of the modulation within this gas remains unclear.\\

Within the sample of detected periods, the 67s period in
Cal\,83 is the longest one, and interpretating this as the
rotation period would imply that the white dwarfs in the
Galactic novae in our sample would be even shorter.
Spinning up to such short rotation rates implies
even longer mass-accretion histories. \cite{osborne11} also
argued that the period changes are not consistent with realistic
torques on the white dwarf. The changes in period by $\pm3$\,s 
argue against rotation of the white dwarf, although
the argument about non-synchronous modulation within
a hydrogen gas envelope presented by \cite{cal83_67sec}
cannot be fully discarded. This non-synchronous region may
be similar to that proposed by \cite{warner04}, who explained
'dwarf nova oscillations' by appealing to weak magnetic fields
that incompletely enforce co-rotation; see also \cite{woudt2010}. 
However, it remains unclear how the rotation period would
manifest itself as an observable period, for instance, whether it
might be caused by periodic absorption patterns.\\

\cite{kteri35} noted that the presence of the same periods
in KT\,Eri and RS\,Oph would appear to argue against a
rotation-based origin for this modulation. Meanwhile, a possible
residual nuclear-burning white dwarf pulsation might be more
constrained in frequency, which would also explain the small range
of detected periods in our sample.\\

The rotation of the white dwarf might be an interpretation
for the observed oscillations if some speculative arguments
were accepted, but we cannot identify under which circumstances
this period can be seen in X-ray observations. In particular we
cannot explain why the signal is transient and why it is seen
in such a small fraction of systems observed. If magnetic fields
are needed, as in intermediate polars, the question arises why
these oscillations are seen only during the SSS phase, while
intermediate polars show the rotation period in
optical and X-ray light curves in quiescence, for example, V4743\,Sgr
\citep{dobrness09}.
An independent measurement of the rotation period for one of
our five systems with short-period oscillations would robustly
confirm or reject the possibility that the rotation period is
the main driver for these oscillations. Meanwhile, other
possibilities such as pulsations exist.

\subsection{Pulsation modes}

An attractive explanation for the observed oscillations is that
it is a surface $g$-mode that is triggered by the $\epsilon$-mechanism.
This possibility gains some favour from considering
the dynamical time-scale of the white dwarf atmosphere. In the
$\epsilon$-mechanism, the contraction
and compression of a mode leads to changes in the nuclear burning
rate. If this energy input exceeds the cooling during the
subsequent rarefaction of the mode, it will be unstably excited.
Since the oscillations are preferentially observed in the SSS following
nova outbursts, when nuclear burning may be occurring at
the surface, the conditions may be met for this to occur.\\

The $\epsilon$-mechanism has been considered for $g$-modes
in the context of white dwarfs following the production of
planetary nebulae \citep{kawaler88}, as well as for neutron stars
that are stably burning their accreted fuel similarly to the
way this occurs in an SSS \citep{piro04}.
Since this scenario is for physical conditions somewhat different from the nova systems studied here, it is helpful to
provide some basic arguments about what types of periods
we expect for surface $g$-modes. This also demonstrates that the mode periods
roughly depend on the properties of the white dwarfs and thus
indicates that these modes may be useful probes of their structure.\\

When the white dwarf is in an SSS state, there is an abrupt
change in the internal profile (in composition, temperature, and density)
where hydrogen is being burned. There is an associated
strong buoyant force at this depth, so that we can consider modes
that are riding along, unable to penetrate this region. If the burning
occurs at a depth $H$, which is much much lower than the radius of the
white dwarf $R$, this mode's dispersion relation will be similar to that
of a shallow water wave \citep{faber95}
\be
\omega^2 \approx g H k^2 \frac{\Delta \rho}{\rho},
\ee
\text{\text{where }}   $ \omega$ is the mode frequency, $g$ is the local gravitational
acceleration, $g\approx GM/R^2$ when $H\ll R$, $k^2=l(l+1)/R^2$
is the transverse wave number squared, where $l$ is the spherical
harmonics quantum number, and $\Delta \rho/\rho$ is the fractional change
in density at the burning depth. For simplicity, we assume $\Delta\rho/\rho\sim1$
in our discussion below, which corresponds to a maximally discontinuous
jump.\\

While $g$ and $k$ depend on the mass and radius of the white dwarf,
$H$ depends on the thermal profile of the surface layers.
The thermal profile of the white dwarf is set via flux-limited
diffusion, which obeys the equation
\be
        \frac{L}{4\pi R^2} = -\frac{4ac T^3}{3\kappa\rho} \frac{\partial T}{\partial r},
\ee
where $L $ is the luminosity, $a$ is the
radiation constant, $\kappa$ is the opacity, $T$ is
the temperature, and $\rho$ is the density. We
assume $\kappa=0.34\,{\rm cm^2\,g^{-1}}$ is a constant as set
by electron scattering for solar composition material, which is roughly
correct for the hot SSS state.
Using hydrostatic balance, $dP/dr= -\rho g$, this expression can
be rewritten as
\be
\frac{L}{4\pi R^2} = \frac{4ac gT^3}{3\kappa} \frac{\partial T}{\partial P}.
\ee
Assuming that $L$ is constant with depth (which is a good approximation
above the burning depth in the SSS),  this can be integrated to
yield
\be
        T(P) = \lp\frac{3P}{a} \rp^{1/4} \lp\frac{L}{L_{\rm Edd}} \rp^{1/4},
\ee
where $L_{\rm Edd} = 4\pi GMc/\kappa$ is the Eddington luminosity.
The layer thickness is then given by
\be
        H \approx \frac{k_{\rm B}T}{\mu m_p g}
        \approx \frac{k_{\rm B}}{\mu m_p g}
        \lp\frac{3P}{a} \rp^{1/4} \lp\frac{L}{L_{\rm Edd}} \rp^{1/4},
\ee
where $k_{\rm B}$ is the Boltzmann constant, $\mu$ is the mean molecular
weight, $m_p$ is the proton mass, and we assume an ideal gas equation
of state. \\

The thickness $H$ therefore depends on the luminosity $L$, which can be
inferred from observations, and the pressure $P$ at the burning depth.
This pressure should be related to $L$ and the requirement of steady burning.
This therefore can be addressed via theoretical calculations.
For now, we know that the pressure must be lower than the pressure at
ignition of the layer \citep[for example, see the calculations of][]{truranlivio86},
so for demonstrational purposes we set
$P\approx10^{18}\,{\rm dyn\,cm^{-2}}$, so that
\begin{equation}
        H\sim2\times10^7\lp\frac{g}{2\times10^9\,{\rm cm\,s^{-2}}} \rp^{-1}
        \lp\frac{P}{10^{18}\,{\rm dyn\,cm^{-2}}} \rp^{1/4}
        \lp\frac{L}{L_{\rm Edd}} \rp^{1/4}{\rm cm},
\end{equation}
where we use $g$ as appropriate for a roughly $\approx1.3M_\odot$, similar
to the kinds of masses inferred for these systems.
Substituting this into the shallow water wave dispersion relation, the
estimated period is
\begin{equation}
        P_{\rm mode} \sim 10\lp\frac{P}{10^{18}\,{\rm dyn\,cm^{-2}}} \rp^{-1/8}
        \lp\frac{L}{L_{\rm Edd}} \rp^{-1/8}
        \lp\frac{R}{3\times10^8\,{\rm cm}} \rp{\rm s},
\end{equation}
where we take $l=1$ since this is the lowest order (and thus easiest to
observe) angular mode, and $\Delta\rho/\rho\sim1$ due to the strong compositional stratification
following the burst. This result highlights that the strongest dependency of $P_{\rm mode}$
is on $R$ and that for typical parameters associated with recurrent novae the period is too low.
The two possible solutions that could reconcile this period with the observed periods
are (1) the white dwarfs are much less massive and have a larger radius, or (2) the
higher order radial modes are excited instead of the lowest order mode we consider here.
The first possibility is unlikely given the constraints on the white dwarf masses from the recurrence
times and nova durations \citep{wolf13}. The second possibility would give a better match
in period since the higher order radial modes increase with the number of radial nodes.
This would then require some explanation for why the higher order mode is excited,
which requires a more detailed theoretical investigation.

If this mode explanation for the oscillations is confirmed, then the period of these modes
provides a constraint on the mass of the white dwarf. This could be useful for determining
whether these nova systems are more massive
than typical white dwarfs, which may indicate that they are growing
from accretion over time. Signs of such mass growth may have
implications for whether some of these systems are attractive as
Type Ia supernova progenitors.

\section{Summary and conclusions}
\label{sect:concl}

We focused on analysing
continuous light curves obtained with \xmm\ and \chandra.
We constructed power spectra that probe the period interval 25--100s
and tested 500 frequencies on an equidistant frequency grid, which corresponds
to an oversampling of 16. We estimated detection probabilities (appendix)
based on simulations of 10,000 synthetic light curves with random
Poisson noise.
We furthermore broke each light curve into 1000s time segments to study
time variability of any periodic signal. We corrected the detection
probabilities for multiple testing by the conservative Bonferroni
correction.\\

We confirm the transient nature of the 35s periods in RS\,Oph
and KT\,Eri and the 54s period in V339\,Del and determined the
fractions of time that the respective periods are detected at
$2\sigma$ significance as listed in Table~\ref{tab:withper}.
The 67s period in Cal\,83 and the 33s period in LMC\,2009a
are detected less frequently.\\

Our dynamic power spectra illustrate variations in power and amplitude
on time scales of fractions of an hour. We also found multiple peaks
in the global power spectra for each observation, suggesting
variations of the period, although the uncertainties in periods
are too large to confirm that. None of these variations correlate
with the X-ray brightness.\\

The amplitudes are found to vary by about $<$15\%.
We checked the possibility that the amplitudes (and thus
detectability) may be degraded by Thomson scattering, which could
explain why so few sources are found with a period and why it is
transient. We found that
the density of the scattering medium would have to exceed
$10^{13}$\,cm$^{-3}$. It depends on the unknown optical depth and
the intrinsic amplitude. The intrinsic amplitude
is probably much lower than 100\%, thus requiring even higher densities, so we
conclude that Thomson scattering has no significant effect.\\

Potential origins of the observed oscillations are the rotation
of the white dwarf or intrinsic pulsations. Only speculative
arguments might explain the observations with rotation of the
white dwarf, and we thus focused on pulsations.
While periods in the 10--100 second range are predicted by
nonradial oscillator modes or pulsations in isolated
white dwarfs, the validity of these models under conditions
of active nuclear burning during a nova outburst is 
difficult to estimate, and new models need to be calculated
to account for the higher luminosity and
the physical conditions of nuclear burning near the
surface. 

\begin{acknowledgements}
A.P. Beardmore and J.P. Osborne 
acknowledge support from the UK Space Agency.
M.H. acknowledges support from an ESA fellowship.
A.L. Piro is supported through NSF grants AST-1205732, PHY-1151197, PHY-1404569, and the Sherman Fairchild Foundation.
A. Dobrotka was supported by the Slovak grant VEGA 1/0511/13.
S. Starrfield gratefully acknowledges partial support
from NSF and NASA grants to ASU.
\end{acknowledgements}

%{\it Facilities:} \facility{XMM-Newton}, \facility{Chandra}.

\bibliographystyle{aa}
\bibliography{cn,jn,rsoph,astron}

\appendix

\section{Appendix}

The period searches were performed using power spectra calculated
with the method of \cite{hornebal}, where the power is
normalised to the total variance of the entire data set.
\cite{scargle82} emphasised that surprisingly large spurious
peaks can occur in a power spectrum and investigated the
probability that a signal
with a given power $z$ is a spurious signal (false-alarm
probability). His Eq.~(14) computes the probability for a
given power peak to be due to a chance noise fluctuation,
including the statistical penalty for inspecting a large
number of frequencies, $N_{\rm f}$. Based on this definition of
the false-alarm probability, $P_{\rm fap}$, we define the
reverse, a detection likelihood ${\cal L}=1-P_{\rm fap}$, thus:
\begin{equation}
\label{psdlike}
{\cal L}=(1-e^{-z})^{N_{\rm f}}
,\end{equation}
from which the power $z$ that yields a given detection likelihood
${\cal L}$ can then be inferred from
\begin{equation}
\label{powlike}
z=-ln(1-{\cal L}^{1/N_{\rm f}})\,.
\end{equation}
For example, in a power spectrum consisting of $N_{\rm f}=500$ tested
frequencies, the required normalised power that yields a 99.9\%
detection probability is $z=13.2$. However, in the discussion that
follows we determine the most appropriate
choice for $N_f$, and hence the detection threshold, in an
oversampled periodogram.\\

In each 1000s time segment studied in this article, the period
range 25--100\,s is probed, corresponding to the frequency interval
0.01--0.04\,Hz and thus a frequency range d$f=0.03$\,Hz.
The sampling above from which no further information can be
gained is to test at the Fourier frequencies, thus with a frequency
bin size of $1/t$, d$f\times t=30$, within the
0.03Hz frequency range. Meanwhile, with $N_{\rm f}=500$ frequencies,
we oversample with a factor 16. By oversampling, we are
more sensitive to periodicities
that might fall midway between the Fourier frequencies.\\

\begin{figure}
 \resizebox{\hsize}{!}{\includegraphics{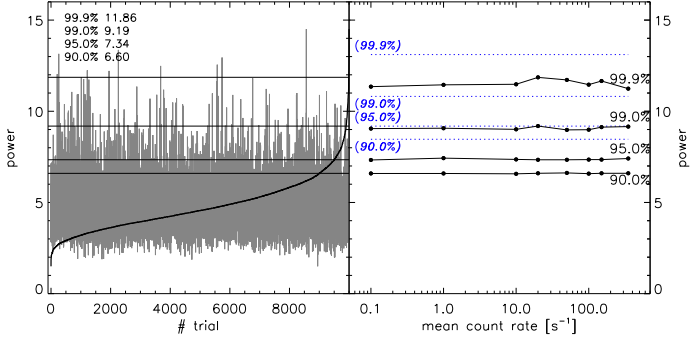}}
\caption{\label{sim_pow}Simulations based on $N=10,000$ synthetic
light curves of 1000s duration with Poisson noise around various given
mean count rates to estimate the likelihood of random detections at
four levels of confidence. Power spectra of the random light curves
were calculated over the period range 25--100\,s, testing 500
frequencies.
{\bf Left panel}: Example of 20 counts per second as
mean count rate. We show the maximum powers in the
order of trial number in grey and sorted by power value with
the thick black line. The sorted curve was used to extract the
power thresholds given in the legend. In the {\bf right panel}
we show the results for the four confidence levels as a function
of mean count rate with connected black bullet points in comparison
to the theoretical thresholds according to Eq.~(\ref{powlike}),
with $N_{\bf f}=500$, included with the dotted blue lines.
}
\end{figure}

As a test of the detection thresholds defined above, we performed
simulations by generating synthetic light curves with random Poisson noise
around a range of given mean count rates. For each given mean count
rate, we generated a sample of 10,000 synthetic light curves of 1000s
duration, the same duration as was chosen for the time segments into which we
split the observations. We then computed power spectra under the same
conditions as for the data, thus probing a period range of 25--100s
with $N_{\rm f}=500$ tested frequencies. Next, we determined the maximum
peak in each power spectrum and show the results in Fig.~\ref{sim_pow}.
In the left panel, the results are shown for the case of a mean count
rate of 20 counts per second. The light grey line indicates the evolution
of random maximum peak powers with trial number, while the thick black
line shows the same data sorted by peak powers. To estimate
the probability of a random encounter at a given confidence level,
the sorted curve can be interpolated. For example, for a detection at 99\%
we determined the power value that corresponds to the sorted curve at
trial number 9000 of the black line, in this case, yielding $z=9.19$.
We extracted the thresholds for four confidence levels and included the
results in the right legend of the left panel for synthetic light
curves with a mean count rate of 20 counts per second. We performed the
same simulations for mean count rates ranging from 0.1--350 counts per
second and
show the results in the right panel of Fig.~\ref{sim_pow}. The simulation
results are shown with the black bullet points, and the predictions by
\cite{scargle82}, reformulated with Eq.~(\ref{powlike}), are included
for the same confidence levels as the blue dotted lines. No systematic
trend with count rate can be identified, but all simulated results
are well below the corresponding theoretically predicted thresholds.\\

\begin{figure}
 \resizebox{\hsize}{!}{\includegraphics{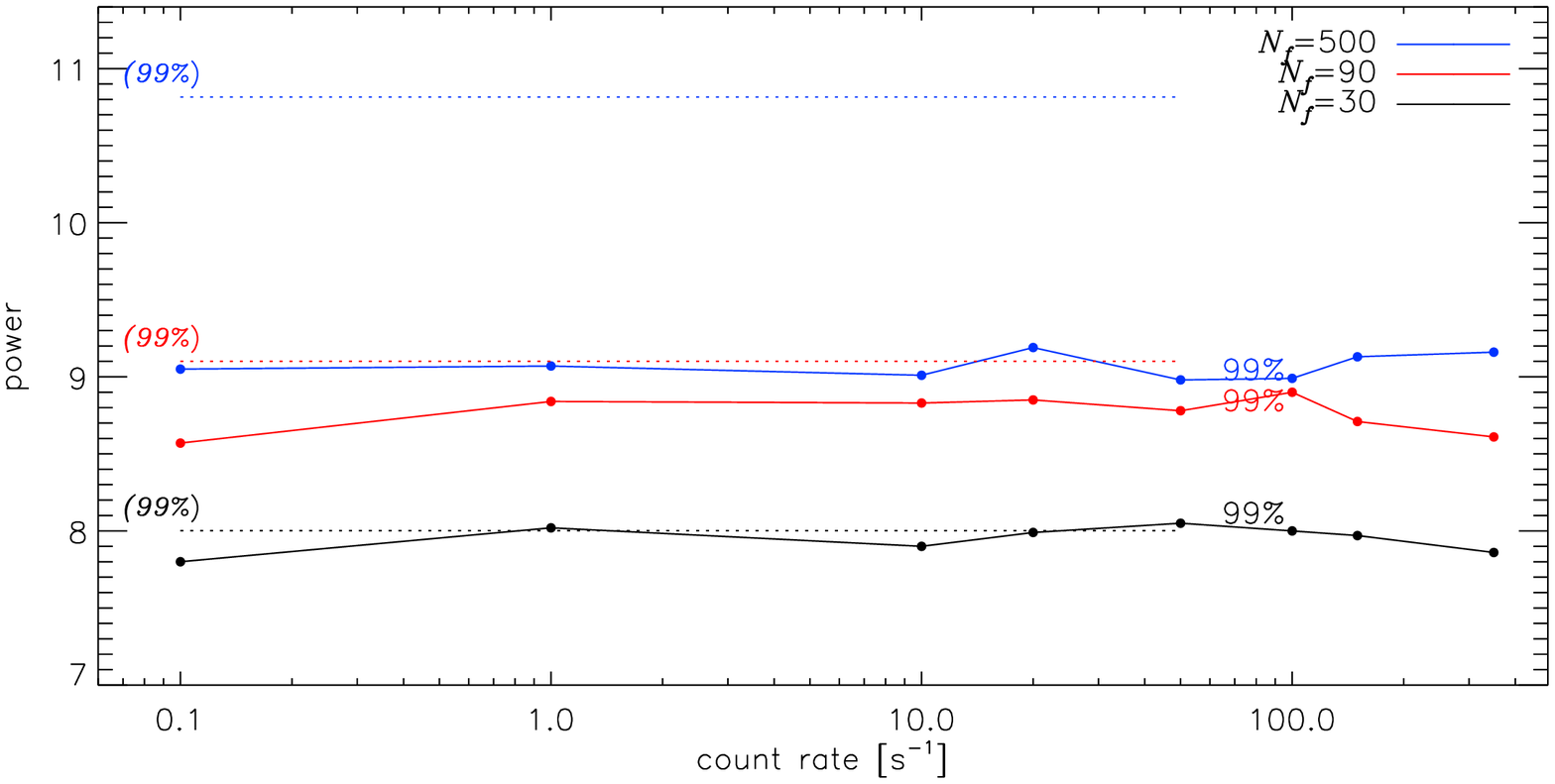}}

 \resizebox{\hsize}{!}{\includegraphics{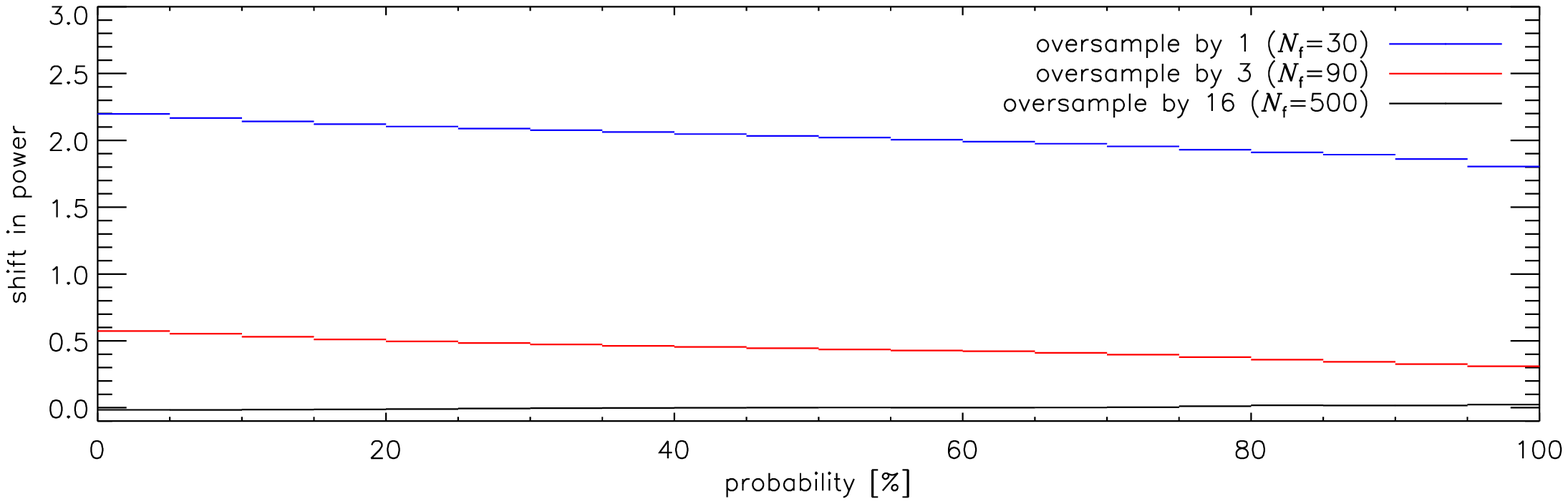}}
\caption{\label{cmp_os}{\bf Top}: Comparison of 99\% detection thresholds
obtained with simulations as in Fig.~\ref{sim_pow} using different
oversampling factors 16 ($N_{\rm f}=500$), 3 ($N_f=90$), and 1
($N_{\rm f}=30$) with the number of tested frequencies $N_f$
given in the legend. The dotted lines are the thresholds
obtained from Eq.~(~\ref{powlike}), and they agree best with
the simulation results without oversampling, while for higher
oversampling rates, the required power for a 99\% detection
is much lower. {\bf Bottom}: shift in power
as a function of detection probability, illustrating
that it is not constant. When determining detection probabilities,
we correct for oversampling by interpolating these simulations.
}
\end{figure}

To understand the discrepancy, we repeated the simulations
with two additional oversampling factors, thus testing 90 frequencies
(oversampling by 3) and 30 frequencies (no oversampling).
In Fig.~\ref{cmp_os} we show the results only for the 99\%
detection thresholds for these three oversampling factors with
black, red, and blue as labelled in the upper right panel. 
In our case, using a period range 25--100s (d$f=0.03$\,Hz) and light
curve duration 1000s, the given values $N_{\rm f}$ need to be divided
by 30 to derive the oversampling factor. The dotted lines are the
thresholds calculated from  $N_{\rm f}$ using  Eq.~(\ref{powlike}),
and the dotted and solid lines agree if no oversampling
is applied. The discrepancies can therefore solely be attributed to
the oversampling. In the bottom panel of Fig.~\ref{cmp_os}, the
difference between powers with and without oversampling is shown for
the same three oversampling factors as a function of detection
probability, from which it can be seen that the discrepancy depends
on the confidence level. Therefore, for this work, we converted
observed power values by interpolating look-up tables computed
from the simulations instead of using Eq.~(\ref{powlike}) to
account for the oversampling effects.\\

An additional effect that needs to be taken into account is
multiple testing when probing a series of time segments. We applied
the conservative Bonferroni correction \citep{bonf36,dunn59,dunn61}
to the detection
thresholds by dividing the p value by the number of time segments.
For example, if we find a single peak in a series of 50 power
spectra and wish to know whether this is significant at 99\%
confidence, we divide the p-value of 0.01 by 50 for the test.
A 99\% detection is thus only achieved if the power value
exceeds a value that corresponds to ${\cal L}=(1-0.01/50)=99.98$\%.\\

\end{document}